\documentclass[twocolumn]{aastex63}

\usepackage{subfigure}
\usepackage{url}
\usepackage{hyperref}
\usepackage{longtable}
\usepackage{natbib}
\usepackage{amsmath}
\usepackage{listings}
\usepackage[normalem]{ulem}
\usepackage{bm}
\usepackage{comment}
\usepackage{array}
\newcolumntype{P}[1]{>{\centering\arraybackslash}p{#1}}
\newcolumntype{M}[1]{>{\centering\arraybackslash}m{#1}}

\usepackage{xcolor, fontawesome}
\definecolor{twitterblue}{RGB}{64,153,255}
\newcommand{\twitter}[1]{\href{https://twitter.com/#1 }{\textcolor{twitterblue}{\faTwitter}\,\tt \textcolor{twitterblue}{@#1}}}

\definecolor{Code}{rgb}{0,0,0}
\definecolor{Decorators}{rgb}{0.5,0.5,0.5}
\definecolor{Numbers}{rgb}{0.5,0,0}
\definecolor{MatchingBrackets}{rgb}{0.25,0.5,0.5}
\definecolor{Keywords}{rgb}{1,0,0}
\definecolor{self}{rgb}{0,0,0}
\definecolor{Strings}{rgb}{0,0.63,0}
\definecolor{Comments}{rgb}{0,0.63,1}
\definecolor{Backquotes}{rgb}{0,0,0}
\definecolor{Classname}{rgb}{0,0,0}
\definecolor{FunctionName}{rgb}{0,0,0}
\definecolor{Operators}{rgb}{0,0,0}
\definecolor{Background}{rgb}{0.98,0.98,0.98}
\definecolor{Booleans}{rgb}{0.572,0,0.572}
\definecolor{BuiltinFunction}{rgb}{0.572,0,0.572}
\definecolor{BuiltinConstant}{rgb}{0.572,0,0.572}
\definecolor{Asterisk}{rgb}{0.670,0,1}

\lstdefinelanguage{Python}{
        numbers=left,
        numberstyle=\footnotesize,
        numbersep=7pt,
        xleftmargin=1.26em,
        framextopmargin=2em,
        framexbottommargin=2em,
        showspaces=false,
        showtabs=false,
        showstringspaces=false,
        frame=l,
        tabsize=4,
        stepnumber=1,
    basicstyle=\small\ttfamily,
        backgroundcolor=\color{Background},
    stringstyle=\ttfamily\color{Strings},
    morekeywords={import,from,class,def,while,if,in,elif,else,not,or,print,break,continue,return,access,as,except,exec,finally,global,import,lambda,pass,print,raise,try,assert},
        keywordstyle={\color{Keywords}\bfseries}, 
    otherkeywords={[2]*},
    keywordstyle={[2]\color{Asterisk}},
}

\usepackage{color}

\newcommand{\shrug}{\texttt{\raisebox{0.75em}{\char`\_}\char`\\\char`\_\kern-0.5ex(\kern-0.25ex\raisebox{0.25ex}{\rotatebox{45}{\raisebox{-.75ex}"\kern-1.5ex\rotatebox{-90})}}\kern-0.5ex)\kern-0.5ex\char`\_/\raisebox{0.75em}{\char`\_}}}

\newcommand{\kep}{{\it Kepler}}

\newcommand{\tess}{{\it TESS}}

\newcommand{\spitz}{{\it Spitzer}}
\newcommand{\vsini}{{$v \sin i$}}

\newcommand{\rstar}{{$R_\star$}}

\newcommand{\rearth}{{R$_\oplus$}}
\newcommand{\rsun}{{R$_\odot$}}

\newcommand{\mjup}{{M$_\textrm{Jup}$}}

\newcommand{\eleanor}{\texttt{eleanor}}

\newcommand{\unsw}{School of Physics, University of New South Wales, Sydney, NSW 2052, Australia}
\newcommand{\chicago}{Department of Astronomy and Astrophysics, University of
Chicago, 5640 S. Ellis Ave, Chicago, IL 60637, USA}

\newcommand{\grfp}{NSF Graduate Research Fellow}

\newcommand{\flatiron}{Center for Computational Astrophysics, Flatiron Institute, 162 Fifth Ave, New York, NY 10010, USA}

\newcommand{\dtm}{Department of Terrestrial Magnetism, Carnegie Institute of Washington, Washington, DC 20015, USA}

\newcommand{\carnegie}{The Observatories of the Carnegie Institution for Science, 813 Santa Barbara St., Pasadena, CA 91101, USA}

\newcommand{\princeton}{Department of Astrophysical Sciences, Princeton University, 4 Ivy Lane, Princeton, NJ 08544, USA}

\newcommand{\baeri}{Bay Area Environmental Research Institute, P.O. Box 25, Moffett Field, CA 94035, USA}

\begin{document}
\title{The Young Planet DS Tuc Ab has a Low Obliquity\footnote{This paper includes data gathered with the 6.5 meter Magellan Telescopes located at Las Campanas Observatory, Chile.}}

\shorttitle{DS Tuc Ab has a Low Obliquity} 
\shortauthors{Montet et al.}

\author[0000-0001-7516-8308]{Benjamin~T.~Montet}
\affiliation{\unsw}
\affiliation{\chicago}

\author[0000-0002-9464-8101]{Adina~D.~Feinstein}
\altaffiliation{\grfp}
\affiliation{\chicago}

\author[0000-0002-0296-3826]{Rodrigo Luger}
\affiliation{\flatiron}

\author[0000-0001-9907-7742]{Megan E. Bedell}
\affiliation{\flatiron}

\author[0000-0002-4020-3457]{Michael A. Gully-Santiago}
\affiliation{\baeri}

\author{Johanna K. Teske}
\altaffiliation{Hubble Fellow}
\affiliation{\carnegie}
\affiliation{\dtm}

\author[0000-0002-6937-9034]{Sharon Xuesong Wang}
\affiliation{\carnegie}
\affiliation{\dtm}

\author{R. Paul Butler}
\affiliation{\dtm}

\author[0000-0001-8045-1765]{Erin Flowers}
\altaffiliation{\grfp}
\affiliation{\princeton}

\author{Stephen A. Shectman}
\affiliation{\carnegie}

\author{Jeffrey D. Crane}
\affiliation{\carnegie}

\author{Ian B. Thompson}
\affiliation{\carnegie}

\correspondingauthor{Benjamin~T.~Montet; \twitter{benmontet}}
\email{b.montet@unsw.edu.au}


\begin{abstract}

The abundance of short-period planetary systems with high orbital obliquities relative to the spin of their host stars is often taken as evidence that scattering processes play important roles in the formation and evolution of these systems.
More recent studies have suggested that wide binary companions can tilt protoplanetary disks, inducing a high stellar obliquity that form through smooth processes like disk migration.
\object{DS Tuc Ab}, a transiting planet with an \deleted{an} 8.138 day period in the 40 Myr Tucana-Horologium association, likely orbits in the same plane as its now-dissipated protoplanetary disk, enabling us to test these theories of disk physics. 
Here, we report on Rossiter-McLaughlin observations of one transit of DS Tuc Ab with the Planet Finder Spectrograph on the Magellan Clay Telescope at Las Campanas Observatory. 
We confirm the previously detected planet by modeling the planet transit and stellar activity signals simultaneously. We test multiple models to describe the stellar activity-induced \added{radial velocity} variations over the night of the transit, finding the obliquity to be low: $\lambda = 12 \pm 13$ degrees, suggesting that this planet likely formed through smooth disk processes and its protoplanetary disk was not significantly torqued by DS Tuc B. The specific stellar activity model chosen affects the results at the $\approx 5$ degree level.
This is the youngest planet to be observed using this technique; we provide a discussion on best practices to accurately measure the observed signal of similar young planets.
\end{abstract}

\keywords{Exoplanets (498), Exoplanet dynamics (490), High resolution spectroscopy (2096), Starspots (1572)}

\section{Introduction} \label{sec:intro}

Each discovered planetary system represents an outcome of the planet formation process, and therefore provides an opportunity to learn about how different planets form in different environments. 
However, each observed present-day system is not a pure laboratory: over billions of years, planet-planet and planet-star gravitational interactions can scatter, torque, migrate, or otherwise perturb orbits, distancing planetary systems from their initial formation state \citep{Kozai62, Lidov62, Fabrycky07, Chatterjee08}. 
This picture is complicated by the fact that in many cases, stellar ages are very poorly known \citep[e.g.][]{Barnes07, Soderblom10}.
These factors make it challenging to develop and test models of planet formation which can explain all observations.

The origins of hot Jupiters are still unclear \citep{Dawson18}.
Kozai-Lidov cycles and tidal friction are often invoked to explain the formation of hot Jupiters \citep{Fabrycky07}, but smooth disk migration provides a reasonable alternative in many cases \citep{Ida08}. 
Multiple channels may be required to explain all of the observed systems: \citet{Nelson17} analyze data from the HAT and WASP exoplanet surveys and find the data can be well-fit by a model in which ${\sim\!85\%}$ of hot Jupiters are formed through high-eccentricity migration and ${\sim\!15\%}$ through disk migration.

A population of low-obliquity planets\footnote{\added{Here and throughout, we refer to obliquity exclusively to describe the relative angle between the spin of the star and the orbit of the planet, not the relation between the spin of the planet and its orbit, which may be detectable for some systems in the near future \citep{Millholland19}.}} is often considered a signature of smooth disk migration \citep{Morton11a, Ford14}. However, \citet{Albrecht12} show that obliquity is an imperfect tracer of the formation history for many stars, as the tidal realignment timescale for a massive, nearby planet can be shorter than the age of the system for many stars with convective outer layers.
\citet{Batygin12} suggests wide binary companions or nearby stars in the birth cluster can torque disks to random inclinations over Myr timescales. In these cases, young planets in binary systems will have random obliquities even at ages of a few Myr, rather than these obliquities being excited by the companion over much slower timescales. \citet{Franchini19} also highlight the possibility that a planet can be tilted out of the plane of the protoplanetary disk by a binary companion in only a few Myr.

Planets in young clusters are valuable resources to provide clean test cases for planet formation. 
Dynamical interactions like the Kozai-Lidov effect can, depending on the system architecture, occur over hundreds of millions or billions of years \citep{Montet15c, Naoz16}. 
For systems with younger ages, we can rule out many slow-timescale dynamical interactions, meaning it is likely that the orbit of the planet traces the orbit of the now-dissipated disk. 
With a statistical sample of the obliquities of young planets, we can test the hypothesis of \citet{Batygin12} to see if the torquing of a disk by a distant perturber is a common process. However, such a survey is limited by the small number of planetary systems around young stars. There are only a handful of transiting planets known to be younger than 100 Myr, identified by the host star's membership in young moving groups or star forming regions \citep{David16, Mann16, David19}.

Recently, Sector 1 data from the Transiting Exoplanet Survey Satellite \citep[\tess, ][]{Ricker14} were used to identify three transits of a planet with an orbital period of 8.14 days around the star \object{DS Tuc A} \citep{Benatti19, Newton19}. 
These papers statistically validated and characterized this planet, \replaced{finding it to have a radius of $5.70 \pm 0.17$ \rearth\ and a transit duration of $0.1324 \pm 0.0005$ days. 
The host star has spectral type G6V, a stellar effective temperature of $5430 \pm 80$ K, and a model-dependent radius of $0.964 \pm 0.029$ or $0.872 \pm 0.027$ \rsun, from \citet{Newton19} and \citet{Benatti19}, respectively.}{finding results broadly consistent with each other. \citet{Newton19} find a planet radius of $5.70 \pm 0.17$ \rearth\ and a transit duration of $0.1324 \pm 0.0005$ days, with the planet orbiting a G6V star with an effective temperature of $5430 \pm 80$ K, and a model-dependent radius of $0.964 \pm 0.029$. \citet{Benatti19} similarly find a planet radius of $5.63 \pm 0.22$ \rearth\ and a transit duration of 0.119 days, with the planet orbiting a G6V star with an effective temperature of $5542 \pm 21$ K, and a model-dependent radius of $0.872 \pm 0.027$ \rsun.}
We refer the reader to those two papers for a detailed discussion of the stellar parameters and transit fits.

DS Tuc is a member of the Tucana-Horologium (Tuc-Hor) association, which has an age of 35-45 Myr \citep{Bell15, Crundall19}. 
DS Tuc itself is a binary, with a K3V companion at a projected separation of 240 AU. 
From \citet{Holman97}, the timescale for Kozai-Lidov interactions is
\begin{equation}
    \tau \approx P_\textrm{planet} \frac{M_\star}{M_\textrm{pert}} \bigg(\frac{a_\textrm{pert}}{a_\textrm{planet}}\bigg)^3 (1-e^2_\textrm{pert})^{3/2},
\label{eq:timescale}
\end{equation}
where $P_\textrm{planet}$ is the the orbital period of a planet with orbital semimajor axis  $a_\textrm{planet}$ about a host of mass $M_\star$, $M_\textrm{pert}$ is the
mass of the perturbing star, and $a_\textrm{pert}$ and $e_\textrm{pert}$ the semimajor axis and eccentricity of the outer object's orbit around the host star/planet system.

From the orbital parameters in \citet{Newton19}, ${M_\star}/{M_\textrm{pert}} \approx 1.2$, ${a_\textrm{pert}}/{a_\textrm{planet}} \approx 2000$, although the posterior distribution is highly skewed to larger values, and $(1-e^2_\textrm{pert})^{3/2} \approx 0.5$, so the timescale $\tau$ from Equation \ref{eq:timescale} is more than 100 Myr, and possibly much longer depending on the true semimajor axis ratio.

As the age of the system is younger than this timescale, the planet likely has not had time to undergo these oscillations and is more likely to trace the orientation of the now-dissipated protoplanetary disk. 
Measuring the obliquity of the stellar spin relative to the orbit of the planet thus enables us to test theories of disk torquing.
We can measure the projected obliquity between the spin of the star and the orbit of the planet through the Rossiter-McLaughlin (R-M) effect, in which an apparent redshift and blueshift in the radial velocity of the star are observed as a transiting planet occults the blueshifted and redshifted hemispheres of the rotating star, respectively \citep{Rossiter24, McLaughlin24}.
DS Tuc Ab is the youngest known planet for which such an observation has been attempted, providing the best laboratory we have to test this theory.
We note that \citet{Zhou19} also obtained three transits of this planet, including two with the Planet Finder Spectrograph (PFS), for a complementary Doppler tomographic analysis of this system. These two works use different data sets and analysis techniques, providing independent checks of the methods and assumptions made in each work.

The rest of this paper is organized as follows:
In Section \ref{sec:obs} we describe the observations.
In Section \ref{sec:analysis} we describe our data analysis.
In Section \ref{sec:results} we present our results.
In Section \ref{sec:discussion}, we discuss best practices for future similar observations of young stars and potential confounding factors, as well as future work.

\section{Observations}
\label{sec:obs}

We obtained data with \deleted{the} PFS on the Magellan Clay Telescope \citep{Crane06, Crane08, Crane10}. 
On 2019 Aug 11 from UT 01:12 to UT 07:10 we obtained 49 \added{radial velocity (RV)} measurements of DS Tuc A. 
The transit duration is 190 minutes, meaning approximately 50\% of the data were obtained in transit while the other half provides information about the out-of-transit RV baseline. Each exposure was 360 seconds in length and was taken with the 0\farcs3 $\times$ 2\farcs5 slit which provides a \added{full-width at half maximum} (FWHM) resolution of $R \approx 130,000$ with 5 pixels per FWHM. All observations were taken with the iodine cell in place \citep{Marcy92}, which imprints a series of narrow lines at known wavelengths to measure the instrument point spread function and wavelength solution at each epoch.

From these spectra, we also derive two spectroscopic activity indicators: the emission flux measured in the Ca \textsc{II} H \& K lines, $S_{\rm HK}$, and the emission in the H$\alpha$ line, ``$S_{\rm H\alpha}$.'' $S_{\rm HK}$ is defined as in \cite{duncan1991} with the updated $R$ continuum area center from \cite{santos2000}, and $S_{\rm H\alpha}$ is defined as in \cite{gomesdasilva2011}.

To characterize the stellar activity-induced variations, we also obtained twelve additional out-of-transit observations of DS Tuc A. These included four observations over two nights on 2019 Aug 21 and 2019 Aug 22, and eight observations over four nights from 2019 Sep 11 to 2019 Sep 14 (all dates UT).
These observations had exposure lengths varying from 360 to 600 seconds under variable sky conditions, with the goal of achieving a similar \added{signal-to-noise ratio} (SNR) in each of these spectra as during the night of the transit, and provide us with the opportunity to measure the RV variability of the star on rotational period timescales.

We also collected a template spectrum of DS Tuc A on the night of the transit, immediately after the observations described above.
This spectrum, obtained under similar conditions and with the same slit but without the iodine cell in the light path, is used in the pipeline RV modeling.
All observations were then analyzed using the standard PFS pipeline \citep{Butler96b}, which divides the spectrum into 2\AA\ chunks and fits each chunk independently.
The resultant RV is the weighted mean of the RVs of each chunk, and the weighted variance across chunks provides an estimate of the uncertainty.
We tested deriving RVs for DS Tuc A using all of the available data and using only data from the night of the transit, finding a lower point-to-point scatter with the latter strategy. 
This is likely due to changes in the line profile shape on rotational timescales as the stellar surface varies.

As the starspots and stellar activity levels change on the surface of the star, this affects the behavior of each 2\AA\ chunk. 
The weight of each chunk is calculated from the behavior of the RVs reported from this particular chunk across the dataset. 
Chunks with more RV information are more likely on average to see a larger effect from these stellar activity changes, leading to these chunks being downweighted as more data are included, increasing the noise in the final output RVs.

As a result, we achieve the highest RV precision for the transit when we only consider data from the night of the transit itself, all collected at approximately the same levels of stellar activity. We thus use the results of two different extractions when considering the data from the night of the transit itself and from the later observations, which typically have approximately twice the RV uncertainty despite generally being observed to the same expected SNR.
This strategy means in practice, there may be a small RV offset between the two subsections of the data. In comparing the two reductions, we find an offset at the 15 m s$^{-1}$ level, but as we only consider the data from the transit night itself when modeling the R-M signal, this should not affect our results.

\begin{figure}[!tbh]
  \begin{center}
    \includegraphics[width=0.5\textwidth, trim={0cm 0.0cm 0cm 0cm}, clip=true]{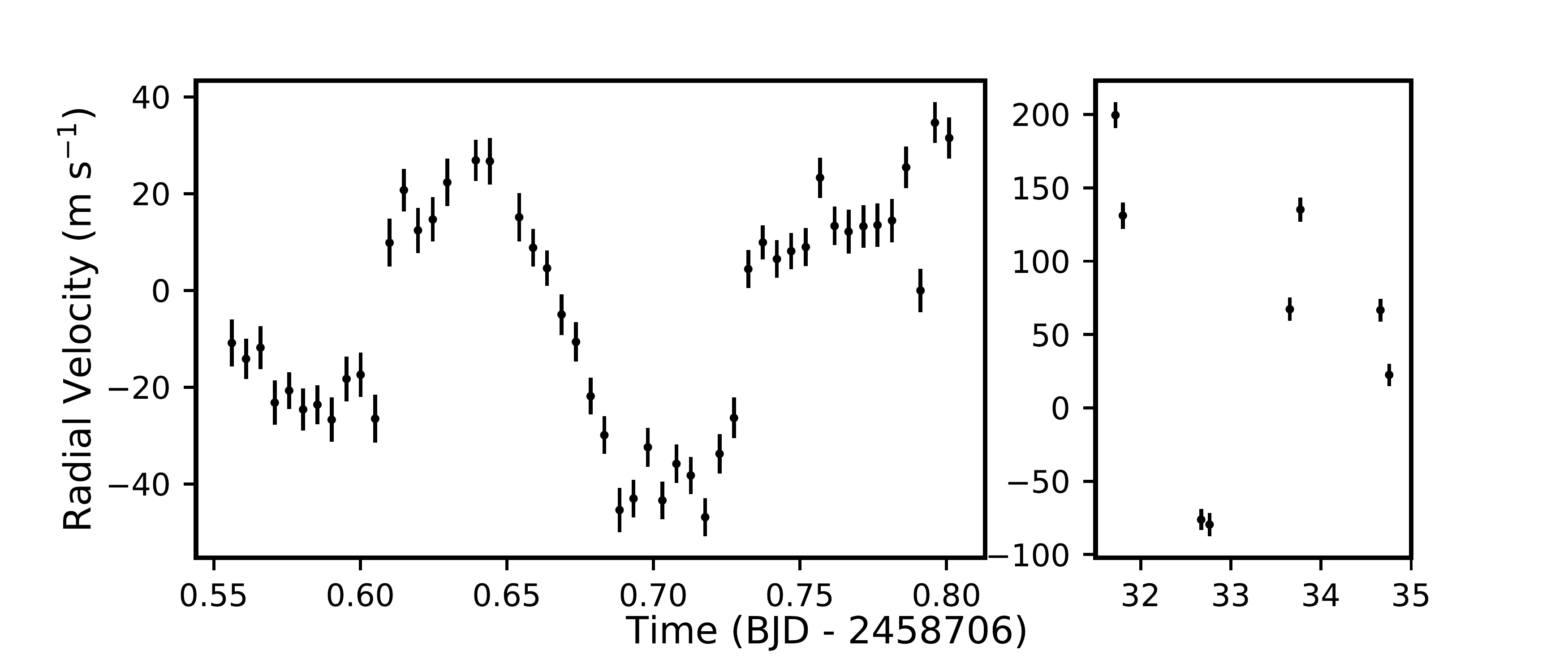}
   \end{center}
  \caption{RV time series for (left) the observed transit and (right) a timespan covering approximately one stellar rotation period, as inferred from \tess\ photometry and discussed in Section \ref{sec:spots}. Note the different vertical scalings on each subplot. The R-M signal is easily detectable, occurring over a significantly different timescale than the rotationally-induced variability signal. The data from the night of the transit and continuing observations were analyzed in two separate data reductions, as described in Section \ref{sec:obs}.}
  \label{fig:data}
\end{figure}

The resultant RVs are given in Table \ref{tab:data} and displayed in Figure \ref{fig:data}.

\section{Data analysis}
\label{sec:analysis}

We model the inferred radial velocities with version 1.0.0.dev3%
\footnote{\url{https://github.com/rodluger/starry/tree/dev}}%
of the \texttt{starry} package of \citet{Luger19}. The \texttt{starry} package models the stellar surface brightness and velocity field as an expansion of spherical harmonics, under which light curve and radial velocity computations are analytic. As the ability to model the Rossiter-McLaughlin effect is currently only available in the development version of the code, we describe it briefly here.

The radial velocity anomaly $\Delta\mathrm{RV}$ due to the Rossiter-McLaughlin may be expressed as \citep[e.g.][]{Gimenez06}
\begin{align}
    \label{eq:DeltaRV}
    \Delta\mathrm{RV} &= \frac{\int I(x, y) v(x, y) \mathrm{d}S}{\int I(x, y)\mathrm{d}S},
\end{align}
where $I(x,y)$ is the intensity at a point $(x, y)$ on the projected disk of the star, $v(x, y)$ is the radial velocity at that point, and the integrals are taken over the visible portion of the projected disk.
The radial component of the velocity field may be written in Cartesian coordinates as \citep[e.g.][]{Short18}
\begin{align}
\label{eq:f_xyz}
v(x, y) = \omega_{eq}(Ax + By)\left(1 - \alpha(-Bx + Ay + Cz)^2\right),
\end{align}
where $z=\sqrt{1 - x^2 - y^2}$, 
\begin{align}
    A &\equiv \sin(i)\cos(\psi) \nonumber \\
    B &\equiv \sin(i)\sin(\psi) \nonumber \\
    C &\equiv \cos(i),
\end{align}
and $i$ and $\psi$ are the stellar inclination and obliquity, respectively. 
The constant $\alpha$ is the linear shear due to differential rotation, whose
effect is to scale the rotational angular velocity of the star according to
\begin{align}
\omega(\theta) = \omega_{eq}(1 - \alpha \sin^2\theta),
\end{align}
where $\omega(\theta)$ is the angular rotational velocity as a function of the latitude $\theta$ and $\omega_{eq}$ is the angular velocity at the equator. 
The coordinate system adopted in Equation~(\ref{eq:f_xyz}) is such that the $x-y$ plane is the plane of the sky and $\hat{z}$ points toward the observer. The planet is assumed to transit the star along the $\hat{x}$ direction in a counterclockwise orbit about $\hat{y}$. The obliquity $\lambda$ is therefore the component of the misalignment angle between the stellar rotation axis and the angular momentum vector of the planet projected onto the plane of the sky.

Because Equation~(\ref{eq:f_xyz}) is a third-degree polynomial in $x$, $y$, and $z$, it may be expressed exactly as a spherical harmonic expansion of degree $l=3$ \citep{Luger19}. This is done by expanding the products in Equation~(\ref{eq:f_xyz}) to obtain the coefficients multiplying each term in the polynomial basis \citep[Equation~7 in][]{Luger19} and transforming them via the spherical harmonic change of basis matrix \citep[Equation~9 in][]{Luger19}.%
\footnote{For a derivation, see \url{https://rodluger.github.io/starry/v1.0.0/notebooks/RossiterMcLaughlin.html}, which is also included as a notebook in the arXiv source material of this manuscript.}%

Under the assumption of quadratic limb darkening, the intensity field $I(x, y)$ may also be expressed exactly as a spherical harmonic expansion of degree $l=2$ \citep{Luger19}. Since the product of two spherical harmonics is also a spherical harmonic, the integrand in Equation~(\ref{eq:DeltaRV}) may be expressed exactly as a degree $l=5$ spherical harmonic expansion. We may therefore use the machinery in \texttt{starry} to \emph{analytically} compute both integrals in Equation~(\ref{eq:DeltaRV}) when the star is occulted by a planet. 
Note that a similar approach was used by \citet{Bedell19}, who model the Rossiter-McLaughlin effect of the hot Jupiter HD\,189733\,b with \texttt{starry}.

Finally, in the sections that follow we explore models in which the star has a single dark Gaussian-shaped spot. In \texttt{starry}, surface features such as spots are also modeled as an expansion of spherical harmonics up to a degree $l_\mathrm{max}$. Again, since spherical harmonics are closed under multiplication, the intensity field $I(x, y)$ of a quadratically limb-darkened spotted surface is simply a sum of spherical harmonics of degree $l_\mathrm{max} + 2$. The total degree of the velocity-weighted intensity (the integrand in Equation~\ref{eq:DeltaRV}) is therefore $l_\mathrm{max} + 5$.

\subsection{Model}
\label{sec:model}

While the R-M signal model is always developed with \texttt{starry} in our analysis, we test multiple approaches to model the effects of the star on these observations.
Over the six hours of the transit, the apparent radial velocity of the star increases by more than 50 m s$^{-1}$. 
In Section \ref{sec:trend} we argue this trend is due to stellar activity rather than an additional unseen planet.
Regardless of the cause of this signal, in order to accurately measure the obliquity of the transiting planet we must model the underlying stellar behavior as well.
We test several different approaches to this problem to ensure our results are not sensitive to our assumptions about the star.
We use low-order polynomials, up to the third degree, to fit the relatively long-term variability during the night of the transit. 
\added{Such low-order polynomials often provide a reasonable description of stellar activity on transit timescales, both in spectroscopic and photometric data \citep[e.g.][]{SanchisOjeda13, NeveuVanMalle16}}.
We also build a stellar activity model by fitting a function that is a linear sum of the observed $S_{\rm H\alpha}$ and calcium $S_{HK}$ measurements during the night, which are correlated with the observed RV, \added{as has been seen in previous analyses of spectroscopic observations of potential planet hosts \citep[e.g.][]{Robertson15, Lanza19}.}
Finally, we also fit \texttt{starry} models of a single starspot in time. The dark spot is modeled as a $l_\mathrm{max} = 4$ spherical harmonic expansion of a symmetric two-dimensional Gaussian flux decrement on the surface of the star; we allow the spot's size, contrast, and location to vary as we fit both signals.

For every model, we can then calculate the expected sum of the R-M signal and the stellar activity signal at each cadence to compare to the data. 
This strategy allows us to understand in a relative sense how well each of the three models fit the data. 
It also gives us the opportunity to verify that the resultant obliquity measurement does not depend on the specific prescription of the stellar activity signal.

\subsection{Likelihood Function}

An accurate likelihood function is critical to ensuring an accurate measurement of the posterior distribution for each parameter.
Each observation has an associated uncertainty, calculated as the weighted standard deviation of the calculated mean of each of the 2\AA\ chunks fit at each epoch.
For this young and active star, this likelihood may not represent the true uncertainty in the measured RV at each epoch. 
For example, occultations of small spots across the surface of the star could cause the observed RV to vary from epoch to epoch, similar to how starspots can affect the observed transit depth in photometric monitoring \citep{SanchisOjeda13}.
Additionally, stellar flares with characteristic timescales similar to a single exposure have been shown to induce RV variations at the 10-100 m s$^{-1}$ level \citep{Reiners09}.

We test multiple likelihood functions with the same form.
First, we assume each data point is drawn from a mixture model \citep[e.g.][]{McLachlan00} which is the sum of two Gaussian functions, each with a different variance:

\begin{equation}
\begin{split}
\mathcal{L} = \prod_i \bigg[ & \frac{q}{\sqrt{2\pi (\sigma^2 + s_1^2)}} \exp\bigg(-\frac{(y_{p,i} + y_{s,i}-V_i)^2}{2(\sigma^2 + s_1^2)}\bigg)  \\+ & \frac{1-q}{\sqrt{2\pi s_2^2}} \exp\bigg(-\frac{(\beta y_{p,i} + y_{s,i}-V_i)^2}{2s_2^2}\bigg) \bigg],
\end{split}
\label{Eq:logl}
\end{equation}
Under this model, the $i$th measurement $V_i$ is compared to our model $y_i$. The model is further subdivided into a contribution from the slowly-varying background RV baseline due to the star, $y_{s,i}$ and from the signal induced by the planet crossing the inhomogeneous surface of the star and blocking some fraction of the stellar disk, $y_{p,i}$. In the definition of this likelihood function, each data point has some probability $q$ of being a ``good'' data point drawn from a relatively narrow distribution. This distribution is a combination of the PFS pipeline uncertainty, $\sigma$, and an additional jitter term, $s_1$, added in quadrature. 
Each data point also has a probability $1-q$ of being drawn from a much broader distribution, with a separate jitter term, $s_2$, to account for the possibility of stellar effects that significantly affect only a small number of data points. We allow the width of both Gaussian distributions to vary in our fitting procedure. 
Additionally, $q$ is a free parameter: if the data were well-modeled by a single Gaussian, we would find the posterior distribution on $q$ to be consistent with 1.0.
Forcing $q=1$ would thus fit the data using only a single Gaussian.

These two Gaussians do not need to have the same mean. 
The surface of the star, as a rapidly rotating G dwarf, is likely dominated by dark starspots rather than bright faculae \citep{Montet17}. 
Each dark spot will induce a signal with roughly the same shape as the R-M effect for an aligned system: as the spot occults the blueshifted hemisphere of the star, it will induce an apparent redshift and vice versa.
As a planet occults a spot then, just as this phenomenon produces a brief brightening in a transit light curve \citep{Desert11, SanchisOjeda13, Morris17}, a spot-crossing event would cause a temporary decrease in the magnitude of the R-M signal. 
To allow for this effect in our fitting procedure, we allow for the broader Gaussian in our mixture model to be offset by some factor which is directly proportional to the magnitude of the R-M signal at that epoch. \added{We parameterize this offset factor in Equation \ref{Eq:logl} as $\beta$.}

When $\beta = 1$, the means of the two Gaussians overlap precisely, leaving us with a standard mixture model as might be expected if the extra variability was not related to starspots. Again, $\beta$ is fit as a free parameter, so if the data were well modeled as a mixture model of two Gaussians with the same mean, we would find the posterior on $\beta$ would be consistent with 1.

We only offset the broader Gaussian. Each starspot is unique, and each draw from our posterior tests only a single $\beta$ value, although disparate starspots with different sizes and different contrast ratios will lead to a different amount of an expected shift from cadence to cadence. 
This uncertainty is manifested through the broader of the two Gaussian terms.
Plainly, this model assumes $q$ is the probability that an observation is not significantly affected by stellar activity. There is then a $1-q$ probability the observation is significantly affected by stellar activity, in which case the magnitude of the effects are relatively uncertain but likely to skew the observation towards smaller absolute values by some factor $\beta$.
We note we also tested models where $\beta$ applied to both Gaussian terms, finding this change did not significantly affect our results.

\subsection{Fitting}

As described in Section \ref{sec:model}, We test three different parameterizations to fit the long-term trend. 
We also test likelihood functions where $\beta$ is fixed at 1.0 and models where $q$ is fixed at 1.0 (in which case $\beta$ is undefined) for each parameterization, giving us nine tests total.

We apply a quadratic limb darkening model for the star, following the prescription of \citet{Kipping13b}. We use uniform priors in both parameters subject to the constraints described in that paper, which enable uninformative sampling of both parameters. We do not fix our limb darkening to theoretical models, which have been shown to induce significant biases for transit observations of planets orbiting active, spotted stars \citep{Csizmadia13}.

We fit these models to evaluate the posterior distribution using the \texttt{emcee} package of \citet{Foreman-Mackey12}, an implementation of the affine-invariant ensemble sampler of \citet{Goodman10}.
For all explorations using the cubic polynomial out-of-transit model, we initialize 500 walkers; for all other runs we initialize 600 walkers. 
We run each for 3,000 steps, removing the first 2,000 as burn-in and considering the final 1,000 steps in our final analysis. 
We verify that our chains have converged following the method of \citet{Geweke92} and through visual inspection.

Here, we assume the planet orbit is circular. 
Given only information about the transit and assuming the orbital velocity of the planet does not change significantly during the transit there is a degeneracy between the eccentricity and the stellar radius. 
We choose to force circular orbits and fit the orbital separation $a/$\rstar\ as a free parameter. The opposite approach would be equally valid and produce similar results.
We include uniform priors on all parameters except for the projected rotational velocity \vsini\ and the radius ratio $R_p$/\rstar.
For both parameters, \citet{Benatti19} and \citet{Newton19} provide discrepant results; we conservatively apply Gaussian priors with means of 18.3 km s$^{-1}$ and 0.057 and standard deviations of 1.8 km s$^{-1}$ and 0.003 for \vsini\ and $R_p$/\rstar, respectively.
\added{In the former case, the two projected rotational velocities are consistent with each other, but the uncertainty on the measurement from \citet{Newton19} is an order of magnitude smaller than the one from \citet{Benatti19}. Here, we use the less precise measurement. 
For $R_p$/\rstar, we choose a value midway between the two results, which have similar published uncertainties, with a width on our prior large enough to encompass both results at $1\sigma$. 
We note that these two parameters both affect the amplitude of the R-M signal but not the asymmetry that signals a projected spin-orbit misalignment. An improper prior on one of these variables would then affect our inference of the other parameter but would not have a significant effect on the measured obliquity.}
While these papers do predict times of transit and impact parameters, we apply uniform priors on each of these parameters as well to allow the possibility that dynamical interactions have affected the transit timing and impact parameter from the \tess\ epochs to the present day.
When applying a spot model directly through \texttt{starry}, we apply uniform priors on the spot latitude, longitude, contrast ratio, and logarithm of the spot size.

\section{Results}
\label{sec:results}

Our results are given in Table \ref{tab:results}. 
The maximum likelihood of the stellar activity indicator model is significantly lower than that of the other models. 
The maximum likelihood $\log \mathcal{L}_\textrm{max}$ differs by only 0.1 between the \texttt{starry} and polynomial models, but is substantially lower for the third model. 
The difference in likelihood corresponds to a Bayes' factor of $\approx 10^{-7}$ when comparing the stellar activity indicator model to the \texttt{starry} model.
In a statistical sense, our starspot model provides approximately an equally valid fit to the data as our simple polynomial model; both provide a much more plausible fit than the stellar activity indicator model.
We note we also tested models using only one of the two stellar activity indicators, but these performed worse than models using both indicators.

This is perhaps not a surprising result: stellar activity indicators are often correlated with RV variations, but not perfectly so \citep{gomesdasilva2011, robertson13}.
As a result, linear models of stellar activity serve as imperfect models to separate planetary and stellar signals, especially on short timescales where the correlation between RVs and stellar activity indicators is even weaker \citep{Meunier19}.
However, because the transit timescale is significantly discrepant from stellar activity timescales and the amplitude of the planetary signal is large, while this model is incomplete it still gives consistent results to our other models. 
This result gives us confidence our results are not likely to be significantly biased by any choices of models that we consider.

All models provide broadly consistent results on the projected obliquity. Considering the two families of most plausible models, the median of the projected obliquity posterior varies from 7 to 14 degrees depending on the specific choice of model used.
All statistical uncertainties range from 11 to 15 degrees.
Using our general polynomial fit, allowing both $q$ and $\beta$ to vary, we infer a projected obliquity of $14 \pm 11$ degrees.
Likewise, with our \texttt{starry} model we infer a projected obliquity of $12 \pm 13$ degrees.
As this model provides the highest likelihood fit to the data we choose this set of values as most representative of our knowledge of the obliquity of the system, but we emphasize that the particular choice of model or likelihood function does not appear to significantly affect the inferred obliquity \added{at the level of more than a few degrees. 
However, the effect of the choice of model is nonzero and fixing a specific model will overestimate the inferred precision of the projected obliquity measurement for this system.}

\begin{figure}[!tbh]
  \begin{center}
    \includegraphics[width=0.45\textwidth, trim={0cm 0.0cm 1cm 1cm}, clip=true]{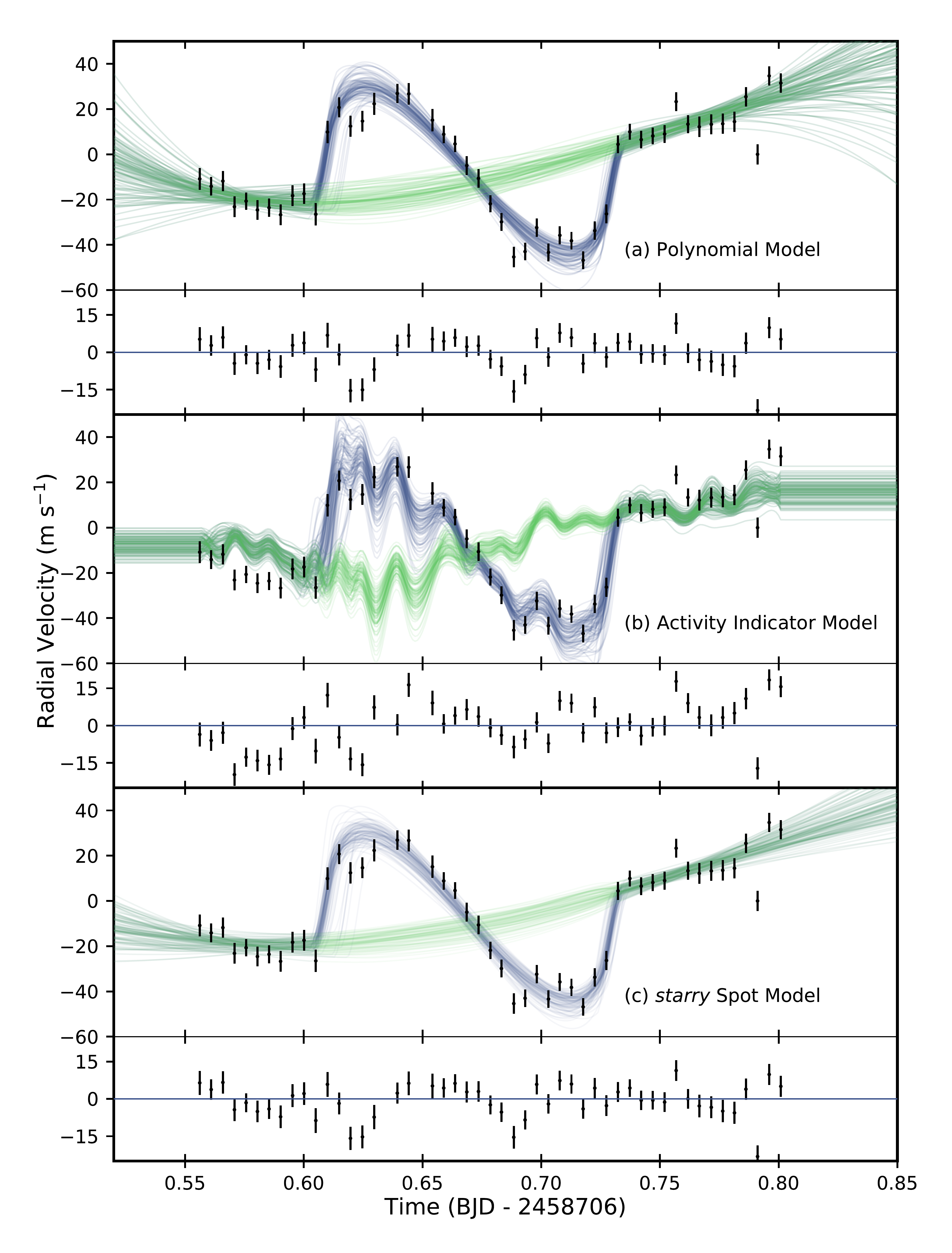}
   \end{center}
  \caption{Draws from the posterior distributions of a simultaneous fit to the
  R-M signal and three different noise models. Green curves represent the noise model, while blue curves include the transit signal as well. From top to bottom, the noise models are a simple cubic polynomial fit, a fit regressed against the spectral stellar activity indicators, and a starspot model built with the \texttt{starry} package of \citep{Luger19}. \added{For each model, the residual to the best-fitting model is shown in the small panel below the posterior draws.} The polynomial and starspot models provide similar quality fits to the data, both of which find significantly higher log likelihood values than the third fit. Significantly, all give consistent results on the projected spin-orbit obliquity angle of approximately $12 \pm 12$ degrees. \added{Correlated residuals during the transit observations are likely from starspot crossing events.} For visualization purposes, we interpolate the stellar activity indicator regression model between the observations with a cubic spline, although the fitting itself only uses information at the times of the observations, where this model is defined.}
  \label{fig:models}
\end{figure}

\added{We plot posterior draws from our fits for each model in Figure \ref{fig:models}, as well as residuals between the data and the best-fitting model. As expected from the likelihood values, visual inspection of the residuals suggest the polynomial model and \texttt{starry} model perform similarly well, and considerably better than the stellar activity model. 
In the residuals, correlated structure can be seen during the transit. This is likely due to starspot crossing events during the transit, as regularly seen in the residuals of transit fits to photometric monitoring of planets orbiting active stars \citep{SanchisOjeda13, Morris17}.}

\citet{Cegla16} provide a relation between the projected obliquity and the true three-dimensional obliquity if the stellar inclination is known. 
The stellar inclination can be inferred from a measurement of the stellar rotation period, \vsini, and stellar radius. 
While there are perhaps significant model uncertainties on the stellar radius, \citet{Newton19} find the stellar inclination consistent with 90 degrees and greater than 70 degrees at 95\% confidence. Assuming these values represent a posterior distribution centered on 90 degrees and with a standard deviation of 10 degrees, we then find the true obliquity \added{$\psi$} between the spin of the host star and orbit of the planet to be less than 16 (27) degrees at $1\sigma$ ($2\sigma$).

\section{Discussion}
\label{sec:discussion}

\subsection{Transit Timing}

The ephemeris from \citet{Newton19}, which includes data from both \tess\ and \spitz, predicts a transit time for this transit of $\textrm{BJD}- 2457000 = 1706.6703 \pm 0.0006$.\footnote{Here, we specifically mean Barycentric Dynamical Time (TDB), which is the time standard used by the \kep\ and \tess\ missions and, unlike Julian Date, is unaffected by leap seconds.}
The ephemeris from \citet{Benatti19}, which includes only \tess\ data, predicts a transit time of  $\textrm{BJD}- 2457000 = 1706.6903 \pm 0.0023$. 
From our data, we measure a transit time of $\textrm{BJD}- 2457000 = 1706.6693 \pm  0.0012$. 
Our result is consistent with the \citet{Newton19} result at the $1\sigma$ level, but is inconsistent with the \citet{Benatti19} result at $8.1\sigma$.
We note that these two predictions, using largely the same original data set, make predictions for the observed epoch which are $8\sigma$ discrepant with each other. 

While it is possible the difference in inferred period between these two previous studies are due to the presence of an additional nearby planet perturbing the transiting planet's period between the observed \tess\ and \spitz\ epochs, it is also possible that one or both analyses underestimated their photometric uncertainty. 
Both analyses used a Gaussian process to model the stellar rotation, but neither directly accounted for the effects of correlated noise due to stellar oscillations.
Standing acoustic waves in the stellar photosphere cause oscillations with a period of about five minutes for Sun-like stars \citep{Deubner75}. These oscillations induce correlated noise in stellar photometric observations which are often non-negligible for stars of a solar mass and radius \citep{Chaplin13}, and may affect the resultant precision in individual transit times inferred from data collected at a two-minute cadence.

Combining the \tess\ and \spitz\ data with our own transit time, we update the orbital period to $P= 8.13825 \pm 0.00003$ days, assuming a linear ephemeris.
We note that we do not explicitly model the p-modes either, and if they contribute significantly to the RV variability, with one data point every seven minutes on average we may be subject to the same underestimation.
We encourage other follow-up measurements of the transit to confirm or refute the presence of transit timing variations in this system.

\subsection{Long-term trend}
\label{sec:trend}

Over the six hours of the transit, the apparent RV of the star increases at approximately 8 m s$^{-1}$ hr$^{-1}$. 
In Section \ref{sec:analysis}, we model this RV shift using a toy model of a single 
starspot group moving across the surface of the star, finding this provides an appropriate fit to the data. 
However, we also use low-order polynomials to attempt to fit the data, finding that simple heuristic works approximately equally well.
This could, in principle, be caused by observing a fraction of a Keplerian orbit from another object orbiting inducing an RV shift on the star. 
We can easily rule out the wide binary companion as the culprit.
From \citet{Liu02}, the RV acceleration from a wide binary companion with mass $m$ at a known separation $\rho$ observed at a distance $d$ is always bounded such that:
\begin{equation}
    \frac{d\textrm{RV}}{dt} < 197 \textrm{m } \textrm{s}^{-1} \textrm{ d}^{-1}
    \frac{m}{M_\odot} \bigg(\frac{d}{1\textrm{pc}}\bigg)^{-2} \bigg(\frac{\rho}{1''}\bigg)^{2}.
\end{equation}
For this companion, the RV trend induced must be no more than 1 m s$^{-1}$ yr$^{-1}$, much less than the observed acceleration.

\citet{Benatti19} show the RV of DS Tuc A is stable at the $\approx 200$ m s$^{-1}$ level on decadal timescales. 
Therefore, if the trend observed during the transit were caused by a planet, it must be a planet with a period shorter than approximately 20 days. 
If this planet is external to DS Tuc Ab, then it must have $m \sin i \geq 10$ \mjup. 
\citet{Benatti19} rule out any such planets in their analysis of the system.
The only plausible planetary companion which could cause this signal but evade detection is a planet in a 1-3 day orbit with a mass of 1-3 times that of Jupiter. 

Such a companion need only be stable for a relatively short time given the age of the system; a companion similar to this one would likely induce significant TTVs which could be detected by continued transit monitoring.
Our additional observations of the system, taken approximately 30 days after the transit and spread over four nights, do indeed show a signal consistent with a $\sim 3$ \mjup\ planet in a 2.8 day orbital period. However, this period is also consistent with the measured stellar rotation period from \tess\ photometry (Figure~\ref{fig:lc_data}), suggesting this periodicity and the long-term behavior we observe during the night of the transit are more likely explained as starspot-induced modulation.

\subsection{Starspot-induced modulation}

The starspot scenario explains the data well from the night of the transit.
The toy model of Section \ref{sec:analysis} demands a starspot on the redshifted hemisphere of the star, rotating away from our line of sight during the transit. 
While we use a single, Gaussian spot in our modeling, in reality spots are non-Gaussian and often appear in groups \citep[e.g.][]{Kilcik11}.

This may explain the excess variability observed in the second half of our transit. 
Over this part of the transit, the observed point-to-point variability is larger, which may be the result of the planet crossing a relatively more inhomogenous hemisphere on observation timescales, causing an increase in the observed variability over this fraction of the observations.

Our median toy model of a single spot, carried over the entire surface of the star, causes in our model a total RV variation of 235 m s$^{-1}$; the 68\% confidence interval on the peak-to-peak RV shift from this starspot ranges from 191 to 299 m s$^{-1}$.
In fact, we observe a $\Delta$RV of 280 m s$^{-1}$ over the four nights of data obtained to trace out a single stellar rotation of DS Tuc A.
Therefore, a single large spot group can explain both the variability observed during the night of the transit itself and the observed RV scatter on rotational timescales.
This does not mean there is only a single spot group on the surface of the star, but rather informs us about the relative asymmetries in spot coverage from hemisphere to hemisphere as the star rotates.

\subsection{Characterizing the Starspots of DS Tuc A}
\label{sec:spots}
The \texttt{starry} model, in addition to matching the observed RV variability, also approximately matches the photometric variability observed during the \tess\ mission. 
This particular spot model induces variability at the $1.9\% \pm 0.4\%$ level at visible wavelengths.

\begin{figure}[!tbh]
  \begin{center}
    \includegraphics[width=0.5\textwidth, trim={0cm 0.0cm 0cm 0cm}, clip=true]{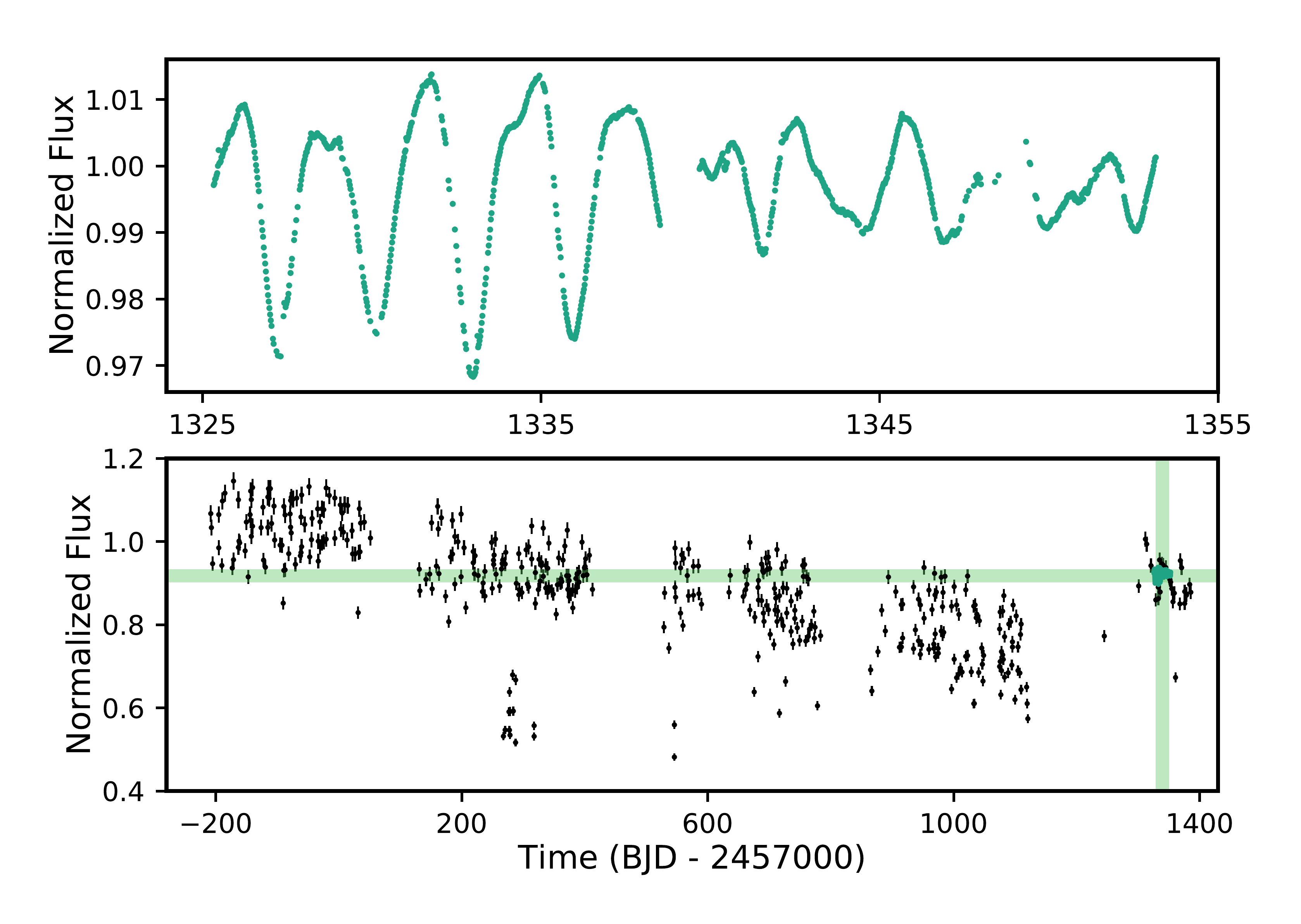}
   \end{center}
  \caption{(Top) DS Tuc A light curve from the \tess\ Sector 1 Full-Frame Images, built with the \eleanor\ pipeline of \citet{Feinstein19}. (Bottom) ASAS-SN light curve for the same star, with the \tess\ light curve overlaid in green. Shaded regions highlight the extent of the \tess\ light curve overlaid on the ASAS-SN data. While \tess\ shows a 4\% variability on rotational timescales, the star itself varies on multi-year timescales by as much as 30\%, suggesting significant starspot coverage on the stellar surface. The brightest observations during the \tess\ observation window are approximately 10\% fainter than the brightest observations with ASAS-SN.}
  \label{fig:lc_data}
\end{figure}

We can compare the modeled spot variability to data from \tess\ itself. 
Figure \ref{fig:lc_data} shows a light curve for DS Tuc A from the \tess\ mission, built using the PSF Flux time series from the \eleanor\ software package of \citet{Feinstein19}. This time series models the point spread function of the detector as a 2D Gaussian at each cadence; the parameters describing the Gaussian are allowed to change from cadence to cadence. 
From these data, a clear rotational signal with a period of $2.85 \pm 0.02$ days can be seen.
It is clear from the \tess\ data that spot groups on the surface of DS Tuc evolve rapidly: at some points in the month of \tess\ data, the variability is at the $\approx$ 4\% level on rotational timescales. 
A few rotation periods later, the spot amplitude is 1\%. 

Therefore, the spot model we use is broadly consistent with not only the observed radial velocity signal, but also the photometric signal.

It is important to note that a single spot or spot group, while appropriate for modeling the asymmetries in starspots which define the observed spectroscopic or photometric modulation, does not represent the entire inhomogeneity of the stellar surface. 
To demonstrate this, Figure \ref{fig:lc_data} also shows for DS Tuc A spanning more than four years from the \added{All-Sky Automated Survey for Supernovae} (ASAS-SN) project \citep{Shappee14, Kochanek17}.
From this light curve, we can see that the overall observed brightness of the star changes by 30\% over four years. 
At the time of the \tess\ data, the star is approximately 10\% fainter than at its 2014 levels, placing a lower limit on the overall spottedness of the star. 
Although the asymmetries in the spot distribution cause photometric modulations at the few percent level, the ASAS-SN data imply that both hemispheres are more spotted than the relative difference in spottedness between the two hemispheres.

\subsection{The Power of Obliquity Measurements of Young Stars}

We have seen from these data that the RV of DS Tuc A varies by nearly 300 m s$^{-1}$ on rotational timescales. 
However, because the surface is relatively consistent on transit timescales, the R-M signal is relatively easy to disentangle from the rotational modulation despite being an order of magnitude smaller in amplitude.

Recent work has shown orbits of planets can still be detected in the face of extreme stellar activity \citep{Barragan19}, and the Doppler method still remains the most effective way to measure masses of planets spectroscopically. 
However, even without a mass we are able to confirm this planet which had only been statistically validated in previous works. 
From the R-M detection, we can be sure this event is the result of an object transiting the surface of DS Tuc A. 
As the transit depth from \tess\ precludes the possibility this event is caused by a transiting brown dwarf or low-mass star, the only possible cause for this observed event is a bona fide transiting planet.
DS Tuc Ab thus joins a small number of planets that have been confirmed through the secure detection of their R-M signal, \added{including Kepler-8b \citep{Jenkins10b} and the Kepler-89/KOI-94 system \citep{Hirano12b}}. DS Tuc Ab is the first member of this class discovered by the \tess\ mission, \added{although future planet candidates discovered around young, active, and rapidly rotating stars discovered as this mission completes its photometric survey of the sky may provide additional opportunities to repeat this procedure.}

DS Tuc A has a measured \vsini~of $18.3 \pm 1.8$ km s$^{-1}$, making precision RV work challenging.
More massive stars that lack convective outer layers can rotate at similar speeds \citep[e.g.][]{McQuillan14}. 
The R-M signal is proportional to the size of the planet and the rotational velocity of the star, and is independent of the mass of either object. 
For rapidly rotating stars or large, low density planets, it is possible to have a larger amplitude R-M signal than Doppler signal. 
Future \tess\ discoveries of young planets or those orbiting massive, rapidly rotating stars may thus find that the best path to confirmation is through observation and characterization of the R-M signal.

For young, rapidly rotating stars, stellar activity can be a limitation to the achievable RV precision.
Large spots can affect the shapes of spectral features, changing the spectrum of the star from observation to observation. 
A template observation of the star may not be representative of the data at another well-separated epoch, limiting the achievable precision.
In this work, we obtained a stellar template on the night of the observations, enabling us to achieve a precision of 3-5 m s$^{-1}$ at most epochs. 
Later observations, although targeted to achieve a similar SNR as our original observations, typically achieve a precision of 7-9 m s$^{-1}$. 
This may be result of line shape variations combined with pipeline systematic effects induced by the interplay between individual iodine and stellar features in the spectra due to a changing barycentric correction between the template and science exposures.
To mitigate these effects for future similar observations of young stars, we encourage future observers to obtain their template spectra, if any are required, as near the transit as feasible so the template reflects a similar state of the stellar surface as the radial velocity observations.

\subsubsection{Differential Rotation}

Differential rotation, in which the equatorial latitudes of a star rotate more quickly than its polar latitudes, can be significant for young stars \citep{Waite17} and therefore may complicate future analyses of the obliquity of young planetary systems.
To quantify the significance of differential rotation, we perform simulations of a rotating star, comparing the observed R-M signal for a star rotating as a solid body to a star with differential rotation such that its equatorial latitudes are rotating with an angular velocity twice that of its polar latitudes. This level of differential rotation is several times larger than the largest values observed for young stars \citep[e.g.][]{Frohlich12}

At its equator, the simulated star has a rotational velocity of 18.3 km s$^{-1}$ and is viewed perfectly edge-on.
We use a simple linear limb darkening model for this experiment, with $u = 0.64$. 
We do not place any spots on the surface of this star.
We then inject a planet the size of Jupiter onto an orbit that transits the surface of this star, measuring the R-M effect from this orbit.

Not surprisingly, for low projected obliquities the effect of differential rotation on the observed signal is small. If the planet crosses the surface of the star occulting a chord of constant latitude, then these data will not sample any of the stellar differential rotation. 
At higher obliquities, as the planet transits regions of the star with different angular velocities at different times, the R-M signal begins to deviate from the uniform rotation case.

\begin{figure}[!tbh]
  \begin{center}
    \includegraphics[width=0.45\textwidth, trim={0cm 0.0cm 0cm 0cm}, clip=true]{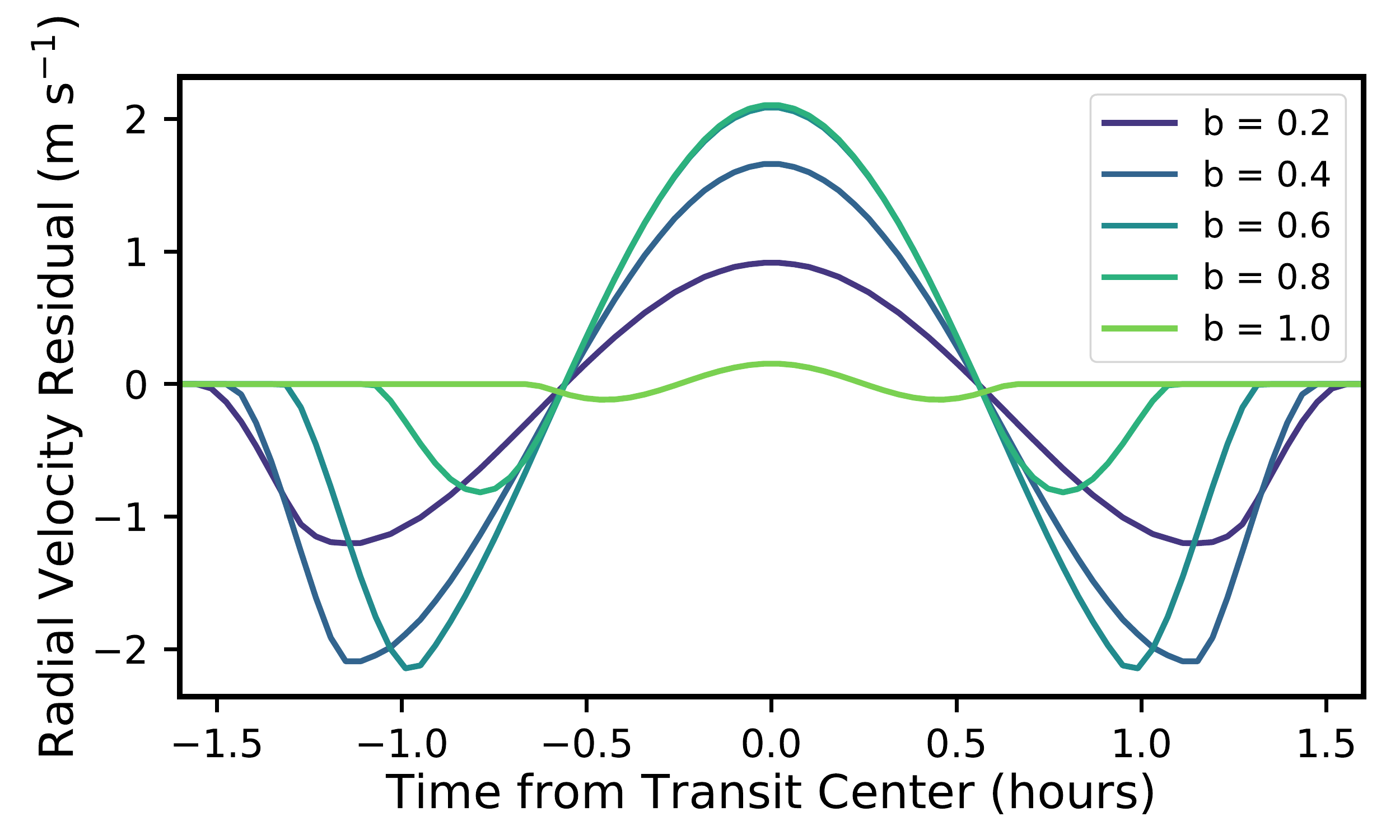}
   \end{center}
  \caption{Difference in best-fitting R-M models for a differentially rotating star and one rotating as a solid body at various impact parameters. At $b=0$ there is no dependence on differential rotation, nor is there for mutually inclined systems. These curves all correspond to $\lambda = 90$ degrees, for a star rotating with an angular velocity twice that at its equator than near its poles and with an equatorial velocity of 18 km s$^{-1}$. Even in this idealized case, the maximum discrepancy between the two models is 2 m s$^{-1}$, so most observations will not be able to detect differential rotation in a single transit. }
  \label{fig:diff_rot}
\end{figure}

Figure \ref{fig:diff_rot} shows the difference between the R-M signal for this star with strong differential rotation and the best-fit signal for a uniformly rotating star at different impact parameters, for a transiting planet with $\lambda = 90$ degrees.
At $b=0$, the planet occults only the central longitudes of the star, where the radial component of the rotational velocity is zero. As a result, the R-M signal for this configuration is zero regardless of differential rotation.
Similarly, at $b=1$ the planet only transits a small range of latitudes, so it does not sample sufficient regions of the star for models of differential or solid body rotation to be distinguished.
However, at intermediate impact parameters, the difference between these models can approach 2 m s$^{-1}$, which in some cases could be detectable. 
In the future, detections of misaligned planets around rapidly rotating stars may provide opportunities to clearly detect differential rotation.

In the case of DS Tuc Ab, its low obliquity means this system is not a viable candidate to detect differential rotation. 
We verify this claim by re-running our spot model fit, allowing the differential rotation shear value $\alpha$ to take any value in the range [0,1] with a uniform prior in that range. 
The resultant posterior is unconstrained: there is power at all values of $\alpha$ and our 95\% confidence interval spans the range [0, 0.88]. 
Significantly, we note including $\alpha$ does not affect our measured projected obliquity: in this run we measure the projected obilquity $\lambda = 12 \pm 14$ degrees, consistent with the $12 \pm 13$ value we find with no differential rotation.

However, flat systems with multiple planets may provide an opportunity to detect differential rotation.
In this work, we infer a rotational velocity of $19.4 \pm 1.5$ km s$^{-1}$ for DS Tuc A.
This is larger than what is measured by \citet{Benatti19} and \citet{Newton19}, albeit only at the $1\sigma$ level.
Observations of two planets orbiting the same star at different impact parameters may provide different inferred rotational velocities, even if both have the same low projected obliquity, if the star has strong differential rotation.
The recently-announced four-planet system V1298 Tau \citep{David19}, a young solar analog with an age of $23 \pm 4$ Myr, may provide such an opportunity in the near future.

\subsubsection{R-M Observations and EPRV pipelines}

The Doppler shift observed through the R-M effect is of course not an intrinsic Doppler shift but rather a change in the shape of the spectral lines. 
The efficacy of this method is reliant on the ability to measure the velocity shift in line centroid through RV processing strategies, which often fit a single, unchanging template to all observations, varying only the velocity shift of this template. 

Previous analyses have tested the ability of different pipelines to produce the expected velocity shifts. For example, \citet{Winn08} and \citet{Johnson08} simulated Keck/HIRES data for HAT-P-1 and TRES-2 and analyzed them with the California Planet Search reduction pipeline, which is very similar to the pipeline used in our DS\,Tuc analysis.
Both these authors found their pipeline recovered the expected transit shifts. \citet{Winn05} and \citet{Hirano10} note the potential discrepancies may be larger for stars with large \vsini, and developed analytic formulae for the analysis of R-M signals, finding the represented accurately the measurements of a cross-correlation analysis for Keck/HIRES and Subaru data, respectively.

More recent work has provided a detailed comparison between R-M and Doppler tomographic methods. \citet{Brown17} compare HARPS spectra of a series of hot Jupiters and measuring their stellar obliquities inferred through the R-M and Doppler tomographic methods. 
They find the results are consistent with each other whether or not the corrections of \citet{Hirano10} are applied, including for stars with \vsini\ values larger than DS Tuc A. There are several other studies of planets orbiting much more rapidly rotating stars which use similar methods \citep[e.g.][]{Triaud09, Gandolfi10}.

We verify our results are not significantly affected by biases induced by the RV pipeline. First, we apply the correction due to this effect produced by \citet{Winn05} for HD 209458 and scaled to the rotational velocity of DS\,Tuc\,A, using our same spot modeling procedure with a single spot as described in Section \ref{sec:model}. In this case, we measure a projected obliquity of $8 \pm 12$ degrees, consistent with that observed previously. 
The $\log\mathcal{L}$ of this model is -162.8, suggesting in this case that the correction of \citet{Winn05} provides a very slightly worse fit than the fit without the correction. As expected, this correction reduces the inferred rotational velocity, from $19.4 \pm 1.5$ km s$^{-1}$ to $17.7 \pm 1.7$ km s$^{-1}$, similar to the effect observed by \citet{Brown17}. 

Additionally, we approximate the PFS pipeline by modeling the line shape variations of a single line transited by a planet. We model the velocity field of a simulated star with a \vsini\ of 18 km s$^{-1}$, following \citet{Short18}, who use this model to derive accurate formulae for the R-M signal. We apply limb darkening parameters consistent with that observed for DS Tuc A. We use this model to build a line profile, then add a Gaussian ``bump'' to represent the transiting planet, following the method of \citet{CollierCameron10}. At each step we infer the velocity change in the centroid of the spectral line, and also cross-correlate with a template model of the spectral line with no transiting planet. We infer the measured RV by fitting a parabola to the cross-correlation result, finding that while there are differences, they are small relative to the size of the signal. The largest discrepancies are at ingress and egress, but importantly are symmetric across the transit, so these discrepancies will not significantly affect the measured obliquity.

This result is not surprising. Standard RV processing pipelines have been shown to produce similar measured obliquities as compared to newer, data-driven approaches with respect to measuring R-M signals \citep{Bedell19}. Discrepancies are generally largest at transit ingress and egress, when the line profile variations are limited to the wings, and our flexible stellar activity model could plausibly account for any short-term systematic offsets.
Moreover, given the measurement of obliquity comes from the potential asymmetry between the blueshifted and redshifted components of the RV curve, any effects from the RV processing pipeline that are symmetric around the RV of the star may affect the measured \vsini\ but should not systematically bias the inferred obliquity.

In general, Doppler tomography analyses and R-M analyses aim to measure the same effect. Doppler tomography, providing a detailed analysis of line profile variations, is feasible with an unstablized spectrograph, while R-M requires a PRV instrument with few m s$^{-1}$ precision. On the other hand, R-M observations provide a more straightforward opportunity to better understand the underlying starspot distribution through the measurement of intranight RV variations. A significant advantage of the R-M method is the development of open-source software built to enable these analyses, while there is no publicly available analysis pipeline for Doppler tomographic analyses. Both methods are able to provide useful results with instruments like PFS, and each provides additional information in some cases that the other cannot provide alone.

\section{Conclusions}

\citet{Batygin12} suggest that wide binary companions may effectively tilt protoplanetary disks, so that a fraction of young, short period planets that migrated through a smooth disk process may nonetheless have high inclinations. 
We look for evidence of this hypothesis by measuring the R-M effect for DS Tuc Ab, a $\approx 40$ Myr planet transiting a Sun-like star in the Tuc-Hor association. 

The orbit of DS Tuc Ab has a low projected obliquity of $\lambda = 12 \pm 13$ degrees relative to the spin of its host star. A single system with a low obliquity neither confirms nor rules out the hypothesis of \citet{Batygin12}, but provides a first data point.
This is the youngest planet for which the R-M effect has been measured. As \tess\ observes more stars that are members of young moving groups, and as data processing techniques with that instrument become more sophisticated, additional planets will be discovered to continue to test this hypothesis.
\added{This result aligns with the conclusions of \citet{Zanazzi18}, who argue that the formation of a warm, giant planet can can reduce or even suppress entirely the excitation of a spin-orbit misalignment, depending on the timescale for accretion onto the planet relative to the disk-binary precession period.
Similar characterization of additional systems like this one may thus be useful in understanding the timescale for giant planet formation in protoplanetary disks.}

\citet{Zhou19} use PFS data of this system for a Doppler tomographic analysis, also finding a low projected obliquity. \citet{Oshagh18} note that changes in the distribution of starspots on highly active stars can affect the measured obliquity by as much as 40 degrees, making additional, complementary observations of this system an important test of the obliquity. 
In this case, observations of multiple transits spread over months enable a confident interpretation of the measurements reported in both \citet{Zhou19} and in this work.

DS Tuc Ab is one of a small number of planets to be confirmed by a detection of its R-M signal rather than its spectroscopic orbit. 
This approach may be the optimal strategy for future confirmation of young planets orbiting rapidly-rotating stars. 
While the RV of the star varies on rotational period timescales at the 300 m s$^{-1}$ level, it does so relatively smoothly over transit timescales, enabling us to cleanly disentangle the stellar and planetary signals.
While this planet would require a dedicated series of many spectra and a detailed data-driven analysis to measure a spectroscopic orbit, the R-M signal is visible by eye in observations from a single night.
For certain systems, in addition to a more amenable noise profile, the amplitude of the R-M signal can be larger than the Doppler amplitude. 
Similar observations to these should be achievable for more young planets as they are discovered, which will shed light onto the end states of planet formation in protoplanetary disks.

\acknowledgements

We thank George Zhou (Center for Astrophysics $|$ Harvard and Smithsonian), Leslie Rogers (University of Chicago), and Chris Spalding (Yale) for conversations which improved the quality of this manuscript.

This paper includes data gathered with the 6.5 meter Magellan Telescopes located at Las Campanas Observatory, Chile.

This research was enabled by the Exostar19 program at the Kavli Institute for Theoretical Physics at UC Santa Barbara, which was supported in part by the National Science Foundation under Grant No. NSF PHY-1748958.

Work by B.T.M. was performed in part under contract with the Jet
Propulsion Laboratory (JPL) funded by NASA through
the Sagan Fellowship Program executed by the NASA
Exoplanet Science Institute.

J.K.T. acknowledges support for this work provided by NASA through Hubble Fellowship grant HST-HF2-51399.001 awarded by the Space Telescope Science Institute, which is operated by the Association of Universities for Research in Astronomy, Inc., for NASA, under contract NAS5-26555.

This material is based upon work supported by the National Science Foundation Graduate Research Fellowship Program under Grant No. DGE-1746045. Any opinions, findings, and conclusions or recommendations expressed in this material are those of the author(s) and do not necessarily reflect the views of the National Science Foundation.

This paper includes data collected by the \tess\ mission. Funding for the \tess\ mission is provided by the NASA Explorer Program.

\tess\ data were obtained from the Mikulski Archive for Space Telescopes
(MAST).
STScI is operated by the Association of Universities for Research in
Astronomy, Inc., under NASA contract NAS5-26555.
Support for MAST is provided by the NASA Office of Space Science via grant
NNX13AC07G and by other grants and contracts.

\software{%
    numpy \citep{numpy},
    matplotlib \citep{matplotlib},
    scipy \citep{Jones01}
    astropy \citep{Astropy18},
    eleanor \citep{Feinstein19},
    starry \citep{Luger19},
    emcee \citep{Foreman-Mackey12}
    }

\facilities{Magellan:Clay (Planet Finder Spectrograph), TESS, ASAS-SN}

\startlongtable
\begin{deluxetable*}{lcccc}
\tablecaption{Derived RVs for DS Tuc A. \label{tab:data}}
\tablehead{
\colhead{Time} & \colhead{RV} & \colhead{Uncertainty} & \colhead{$S_{HK}$} & \colhead{H$\alpha$} \\
\colhead{(BJD)} & \colhead{(m s$^{-1}$)} & \colhead{(m s$^{-1}$)} & {} & {}
}
\startdata
2458706.55618 &  -10.84 & 4.83 &  0.5826 & 0.05566 \\
2458706.56101 &  -14.13 & 4.14 &  0.5735 & 0.05596 \\
2458706.56592 &  -11.80 & 4.45 &  0.5679 & 0.05615 \\
2458706.57080 &  -23.17 & 4.59 &  0.5712 & 0.05566 \\
2458706.57570 &  -20.68 & 3.82 &  0.5727 & 0.05597 \\
2458706.58047 &  -24.58 & 4.33 &  0.5771 & 0.05603 \\
2458706.58537 &  -23.59 & 4.01 &  0.5737 & 0.05593 \\
2458706.59023 &  -26.70 & 4.58 &  0.5809 & 0.05613 \\
2458706.59529 &  -18.26 & 4.61 &  0.5849 & 0.05630 \\
2458706.60010 &  -17.40 & 4.56 &  0.5887 & 0.05648 \\
2458706.60506 &  -26.49 & 4.97 &  0.5993 & 0.05594 \\
2458706.60996 &    9.87 & 4.96 &  0.5900 & 0.05670 \\
2458706.61486 &   20.74 & 4.41 &  0.5936 & 0.05597 \\
2458706.61966 &   12.43 & 4.68 &  0.5992 & 0.05641 \\
2458706.62472 &   14.68 & 4.59 &  0.5853 & 0.05667 \\
2458706.62965 &   22.34 & 4.90 &  0.6046 & 0.05718 \\
2458706.63939 &   26.89 & 4.26 &  0.5839 & 0.05625 \\
2458706.64420 &   26.72 & 4.79 &  0.5985 & 0.05671 \\
2458706.65419 &   15.13 & 4.97 &  0.5898 & 0.05628 \\
2458706.65893 &    8.84 & 3.88 &  0.5716 & 0.05611 \\
2458706.66370 &    4.60 & 3.64 &  0.5712 & 0.05614 \\
2458706.66866 &   -4.96 & 4.21 &  0.5858 & 0.05610 \\
2458706.67356 &  -10.62 & 4.09 &  0.5781 & 0.05598 \\
2458706.67856 &  -21.84 & 3.79 &  0.5755 & 0.05600 \\
2458706.68325 &  -29.88 & 3.91 &  0.5752 & 0.05585 \\
2458706.68844 &  -45.36 & 4.55 &  0.5793 & 0.05599 \\
2458706.69319 &  -42.99 & 3.89 &  0.5684 & 0.05592 \\
2458706.69809 &  -32.38 & 4.02 &  0.5645 & 0.05529 \\
2458706.70304 &  -43.37 & 3.90 &  0.5606 & 0.05518 \\
2458706.70783 &  -35.81 & 3.98 &  0.5680 & 0.05540 \\
2458706.71273 &  -38.21 & 3.85 &  0.5670 & 0.05539 \\
2458706.71765 &  -46.83 & 3.91 &  0.5659 & 0.05522 \\
2458706.72256 &  -33.74 & 4.06 &  0.5646 & 0.05535 \\
2458706.72746 &  -26.33 & 4.21 &  0.5659 & 0.05535 \\
2458706.73235 &    4.43 & 3.97 &  0.5588 & 0.05521 \\
2458706.73731 &    9.93 & 3.52 &  0.5567 & 0.05513 \\
2458706.74210 &    6.50 & 3.89 &  0.5596 & 0.05489 \\
2458706.74700 &    8.13 & 3.74 &  0.5571 & 0.05511 \\
2458706.75197 &    8.98 & 3.97 &  0.5574 & 0.05508 \\
2458706.75683 &   23.30 & 4.15 &  0.5646 & 0.05516 \\
2458706.76179 &   13.34 & 3.98 &  0.5647 & 0.05524 \\
2458706.76657 &   12.16 & 4.56 &  0.5607 & 0.05499 \\
2458706.77163 &   13.26 & 4.39 &  0.5632 & 0.05460 \\
2458706.77643 &   13.52 & 4.49 &  0.5623 & 0.05483 \\
2458706.78139 &   14.45 & 4.47 &  0.5607 & 0.05495 \\
2458706.78620 &   25.45 & 4.27 &  0.5590 & 0.05460 \\
2458706.79111 &    0.00 & 4.46 &  0.5577 & 0.05445 \\
2458706.79602 &   34.67 & 4.21 &  0.5524 & 0.05464 \\
2458706.80090 &   31.51 & 4.25 &  0.5585 & 0.05452 \\ 
\hline
2458716.59155 &  100.30 & 5.84 & -- & -- \\
2458716.72215 &  105.19 & 6.25 & -- & -- \\
2458717.62611 &  146.51 & 8.11 & -- & -- \\
2458717.77760 &  -71.01 & 7.04 & -- & -- \\
2458737.72011 &  199.55 & 8.73 & -- & -- \\
2458737.80174 &  131.10 & 8.98 & -- & -- \\
2458738.67055 &  -76.18 & 7.18 & -- & -- \\
2458738.76341 &  -79.52 & 7.87 & -- & -- \\
2458739.65311 &   67.24 & 7.91 & -- & -- \\
2458739.77091 &  135.17 & 8.23 & -- & -- \\
2458740.65811 &   66.63 & 7.77 & -- & -- \\
2458740.75524 &   22.49 & 7.56 & -- & -- \\
\enddata
\end{deluxetable*}
\begin{deluxetable*}{lccc}[!ht]
\tablecaption{Inferred transit parameters \label{tab:results}}
\tablehead{
\colhead{} & \colhead{Polynomial Model} & \colhead{Stellar Activity Correlation Model} & \colhead{Starspot Fit} 
}
\startdata
\vsini\ (km s$^{-1}$) & $19.6 \pm 1.5$ & $20.0 \pm 1.6$ & $19.4 \pm 1.5$ \\
$R_p/R_\star$ & $0.059 \pm 0.002$ & $0.060 \pm 0.003$ & $0.059 \pm 0.002$ \\
$t_0$ ($\textrm{BJD}-2457000$) & $1706.6692 \pm 0.0010$ & $1706.6691 \pm 0.0018$ &
              $1706.6693 \pm 0.0012$\\
$b$ & $0.18 \pm 0.11$ & $0.17 \pm 0.13$ & $0.18 \pm 0.12$ \\
$a/$\rstar & $20.8 \pm 0.7$ & $21.2 \pm 1.1$ & $20.9 \pm 0.8$ \\
Obliquity (deg) & $14 \pm 11$ & $5 \pm 11$ & $12 \pm 13$ \\
Obliquity (deg), $q=1$ & $12 \pm 11$ & $13 \pm 6$ & $7 \pm 12$ \\
Obliquity (deg), $\alpha = 1$ & $14 \pm 13$ & $3 \pm 11$ & $8 \pm 15$ \\
\hline
Spot Amplitude & - & - & $0.019 \pm 0.005$ \\
Spot Size (\rstar) & - & - & $0.055 \pm 0.023$ \\
Spot Longitude\tablenotemark{a} (deg) & - & - & $26 \pm 4$ \\
Spot Latitude\tablenotemark{a} (deg) & - & - & $28 \pm 8$ \\
\hline
jitter 1 (m s$^{-1}$) & $1.8 \pm 0.9$ & $1.5 \pm 0.9$ & $1.8 \pm 0.9$ \\
jitter 2 (m s$^{-1}$) & $8.8 \pm 4.4$ & $11.9 \pm 3.0$ & $9.3 \pm 5.4$ \\
$q$ & $0.54 \pm 0.24$ & $0.32 \pm 0.17$ & $0.58 \pm 0.24$ \\
$\beta$ & $0.88 \pm 0.08$ & $0.88 \pm 0.11$ & $0.85 \pm 0.15$ \\
\hline
$\log \mathcal{L}_\textrm{max}$ & $-162.2$ & $-178.3$ & $-162.1$ \\
Bayes' Factor & 0.83 & $9.1 \times 10^{-8} $ & 1.0
\enddata
\tablenotetext{a}{Defined at $\textrm{BJD}-2457000 = 1706.5$}
\end{deluxetable*}


\begin{thebibliography}{}
\expandafter\ifx\csname natexlab\endcsname\relax\def\natexlab#1{#1}\fi
\providecommand{\url}[1]{\href{#1}{#1}}

\bibitem[{{Albrecht} {et~al.}(2012){Albrecht}, {Winn}, {Johnson}, {Howard},
  {Marcy}, {Butler}, {Arriagada}, {Crane}, {Shectman}, {Thompson}, {Hirano},
  {Bakos}, \& {Hartman}}]{Albrecht12}
{Albrecht}, S., {Winn}, J.~N., {Johnson}, J.~A., {et~al.} 2012, \apj, 757, 18

\bibitem[{{Astropy Collaboration} {et~al.}(2018){Astropy Collaboration},
  {Price-Whelan}, {Sip{\H o}cz}, {G{\"u}nther}, {Lim}, {Crawford}, {Conseil},
  {Shupe}, {Craig}, {Dencheva}, {Ginsburg}, {VanderPlas}, {Bradley},
  {P{\'e}rez-Su{\'a}rez}, {de Val-Borro}, {Aldcroft}, {Cruz}, {Robitaille},
  {Tollerud}, {Ardelean}, {Babej}, {Bach}, {Bachetti}, {Bakanov}, {Bamford},
  {Barentsen}, {Barmby}, {Baumbach}, {Berry}, {Biscani}, {Boquien}, {Bostroem},
  {Bouma}, {Brammer}, {Bray}, {Breytenbach}, {Buddelmeijer}, {Burke},
  {Calderone}, {Cano Rodr{\'{\i}}guez}, {Cara}, {Cardoso}, {Cheedella},
  {Copin}, {Corrales}, {Crichton}, {D'Avella}, {Deil}, {Depagne}, {Dietrich},
  {Donath}, {Droettboom}, {Earl}, {Erben}, {Fabbro}, {Ferreira}, {Finethy},
  {Fox}, {Garrison}, {Gibbons}, {Goldstein}, {Gommers}, {Greco}, {Greenfield},
  {Groener}, {Grollier}, {Hagen}, {Hirst}, {Homeier}, {Horton}, {Hosseinzadeh},
  {Hu}, {Hunkeler}, {Ivezi{\'c}}, {Jain}, {Jenness}, {Kanarek}, {Kendrew},
  {Kern}, {Kerzendorf}, {Khvalko}, {King}, {Kirkby}, {Kulkarni}, {Kumar},
  {Lee}, {Lenz}, {Littlefair}, {Ma}, {Macleod}, {Mastropietro}, {McCully},
  {Montagnac}, {Morris}, {Mueller}, {Mumford}, {Muna}, {Murphy}, {Nelson},
  {Nguyen}, {Ninan}, {N{\"o}the}, {Ogaz}, {Oh}, {Parejko}, {Parley}, {Pascual},
  {Patil}, {Patil}, {Plunkett}, {Prochaska}, {Rastogi}, {Reddy Janga},
  {Sabater}, {Sakurikar}, {Seifert}, {Sherbert}, {Sherwood-Taylor}, {Shih},
  {Sick}, {Silbiger}, {Singanamalla}, {Singer}, {Sladen}, {Sooley},
  {Sornarajah}, {Streicher}, {Teuben}, {Thomas}, {Tremblay}, {Turner},
  {Terr{\'o}n}, {van Kerkwijk}, {de la Vega}, {Watkins}, {Weaver}, {Whitmore},
  {Woillez}, {Zabalza}, \& {Astropy Contributors}}]{Astropy18}
{Astropy Collaboration}, {Price-Whelan}, A.~M., {Sip{\H o}cz}, B.~M., {et~al.}
  2018, \aj, 156, 123

\bibitem[{{Barnes}(2007)}]{Barnes07}
{Barnes}, S.~A. 2007, \apj, 669, 1167

\bibitem[{{Barrag{\'a}n} {et~al.}(2019){Barrag{\'a}n}, {Aigrain}, {Kubyshkina},
  {Gand olfi}, {Livingston}, {Fridlund}, {Fossati}, {Korth}, {Parviainen},
  {Malavolta}, {Palle}, {Deeg}, {Nowak}, {Rajpaul}, {Zicher}, {Antoniciello},
  {Narita}, {Albrecht}, {Bedin}, {Cabrera}, {Cochran}, {de Leon},
  {Eigm{\"u}ller}, {Fukui}, {Granata}, {Grziwa}, {Guenther}, {Hatzes},
  {Kusakabe}, {Latham}, {Libralato}, {Luque},
  {Monta{\~n}{\'e}s-Rodr{\'\i}guez}, {Murgas}, {Nardiello}, {Pagano}, {Piotto},
  {Persson}, {Redfield}, \& {Tamura}}]{Barragan19}
{Barrag{\'a}n}, O., {Aigrain}, S., {Kubyshkina}, D., {et~al.} 2019, \mnras,
  490, 698

\bibitem[{{Batygin}(2012)}]{Batygin12}
{Batygin}, K. 2012, \nat, 491, 418

\bibitem[{{Bedell} {et~al.}(2019){Bedell}, {Hogg}, {Foreman-Mackey}, {Montet},
  \& {Luger}}]{Bedell19}
{Bedell}, M., {Hogg}, D.~W., {Foreman-Mackey}, D., {Montet}, B.~T., \& {Luger},
  R. 2019, \aj, 158, 164

\bibitem[{{Bell} {et~al.}(2015){Bell}, {Mamajek}, \& {Naylor}}]{Bell15}
{Bell}, C.~P.~M., {Mamajek}, E.~E., \& {Naylor}, T. 2015, \mnras, 454, 593

\bibitem[{{Benatti} {et~al.}(2019){Benatti}, {Nardiello}, {Malavolta},
  {Desidera}, {Borsato}, {Nascimbeni}, {Damasso}, {D'Orazi}, {Mesa}, {Messina},
  {Esposito}, {Bignamini}, {Claudi}, {Covino}, {Lovis}, \&
  {Sabotta}}]{Benatti19}
{Benatti}, S., {Nardiello}, D., {Malavolta}, L., {et~al.} 2019, \aap, 630, A81

\bibitem[{{Brown} {et~al.}(2017){Brown}, {Triaud}, {Doyle}, {Gillon}, {Lendl},
  {Anderson}, {Collier Cameron}, {H{\'e}brard}, {Hellier}, {Lovis}, {Maxted},
  {Pepe}, {Pollacco}, {Queloz}, \& {Smalley}}]{Brown17}
{Brown}, D.~J.~A., {Triaud}, A.~H.~M.~J., {Doyle}, A.~P., {et~al.} 2017,
  \mnras, 464, 810

\bibitem[{{Butler} {et~al.}(1996){Butler}, {Marcy}, {Williams}, {McCarthy},
  {Dosanjh}, \& {Vogt}}]{Butler96b}
{Butler}, R.~P., {Marcy}, G.~W., {Williams}, E., {et~al.} 1996, \pasp, 108, 500

\bibitem[{{Cegla} {et~al.}(2016){Cegla}, {Lovis}, {Bourrier}, {Beeck},
  {Watson}, \& {Pepe}}]{Cegla16}
{Cegla}, H.~M., {Lovis}, C., {Bourrier}, V., {et~al.} 2016, \aap, 588, A127

\bibitem[{{Chaplin} \& {Miglio}(2013)}]{Chaplin13}
{Chaplin}, W.~J., \& {Miglio}, A. 2013, \araa, 51, 353

\bibitem[{{Chatterjee} {et~al.}(2008){Chatterjee}, {Ford}, {Matsumura}, \&
  {Rasio}}]{Chatterjee08}
{Chatterjee}, S., {Ford}, E.~B., {Matsumura}, S., \& {Rasio}, F.~A. 2008, \apj,
  686, 580

\bibitem[{{Collier Cameron} {et~al.}(2010){Collier Cameron}, {Bruce}, {Miller},
  {Triaud}, \& {Queloz}}]{CollierCameron10}
{Collier Cameron}, A., {Bruce}, V.~A., {Miller}, G.~R.~M., {Triaud},
  A.~H.~M.~J., \& {Queloz}, D. 2010, \mnras, 403, 151

\bibitem[{{Crane} {et~al.}(2006){Crane}, {Shectman}, \& {Butler}}]{Crane06}
{Crane}, J.~D., {Shectman}, S.~A., \& {Butler}, R.~P. 2006, in \procspie, Vol.
  6269, 626931

\bibitem[{{Crane} {et~al.}(2010){Crane}, {Shectman}, {Butler}, {Thompson},
  {Birk}, {Jones}, \& {Burley}}]{Crane10}
{Crane}, J.~D., {Shectman}, S.~A., {Butler}, R.~P., {et~al.} 2010, in Society
  of Photo-Optical Instrumentation Engineers (SPIE) Conference Series, Vol.
  7735, \procspie, 773553

\bibitem[{{Crane} {et~al.}(2008){Crane}, {Shectman}, {Butler}, {Thompson}, \&
  {Burley}}]{Crane08}
{Crane}, J.~D., {Shectman}, S.~A., {Butler}, R.~P., {Thompson}, I.~B., \&
  {Burley}, G.~S. 2008, in Society of Photo-Optical Instrumentation Engineers
  (SPIE) Conference Series, Vol. 7014, \procspie, 701479

\bibitem[{{Crundall} {et~al.}(2019){Crundall}, {Ireland}, {Krumholz},
  {Federrath}, {{\v{Z}}erjal}, \& {Hansen}}]{Crundall19}
{Crundall}, T.~D., {Ireland}, M.~J., {Krumholz}, M.~R., {et~al.} 2019, \mnras,
  489, 3625

\bibitem[{{Csizmadia} {et~al.}(2013){Csizmadia}, {Pasternacki}, {Dreyer},
  {Cabrera}, {Erikson}, \& {Rauer}}]{Csizmadia13}
{Csizmadia}, S., {Pasternacki}, T., {Dreyer}, C., {et~al.} 2013, \aap, 549, A9

\bibitem[{{David} {et~al.}(2019){David}, {Petigura}, {Luger}, {Foreman-Mackey},
  {Livingston}, {Mamajek}, \& {Hillenbrand}}]{David19}
{David}, T.~J., {Petigura}, E.~A., {Luger}, R., {et~al.} 2019, \apjl, 885, L12

\bibitem[{{David} {et~al.}(2016){David}, {Hillenbrand}, {Petigura},
  {Carpenter}, {Crossfield}, {Hinkley}, {Ciardi}, {Howard}, {Isaacson}, {Cody},
  {Schlieder}, {Beichman}, \& {Barenfeld}}]{David16}
{David}, T.~J., {Hillenbrand}, L.~A., {Petigura}, E.~A., {et~al.} 2016, \nat,
  534, 658

\bibitem[{{Dawson} \& {Johnson}(2018)}]{Dawson18}
{Dawson}, R.~I., \& {Johnson}, J.~A. 2018, \araa, 56, 175

\bibitem[{{D{\'e}sert} {et~al.}(2011){D{\'e}sert}, {Charbonneau}, {Demory},
  {Ballard}, {Carter}, {Fortney}, {Cochran}, {Endl}, {Quinn}, {Isaacson},
  {Fressin}, {Buchhave}, {Latham}, {Knutson}, {Bryson}, {Torres}, {Rowe},
  {Batalha}, {Borucki}, {Brown}, {Caldwell}, {Christiansen}, {Deming},
  {Fabrycky}, {Ford}, {Gilliland}, {Gillon}, {Haas}, {Jenkins}, {Kinemuchi},
  {Koch}, {Lissauer}, {Lucas}, {Mullally}, {MacQueen}, {Marcy}, {Sasselov},
  {Seager}, {Still}, {Tenenbaum}, {Uddin}, \& {Winn}}]{Desert11}
{D{\'e}sert}, J.-M., {Charbonneau}, D., {Demory}, B.-O., {et~al.} 2011, \apjs,
  197, 14

\bibitem[{{Deubner}(1975)}]{Deubner75}
{Deubner}, F.~L. 1975, \aap, 44, 371

\bibitem[{{Duncan} {et~al.}(1991){Duncan}, {Vaughan}, {Wilson}, {Preston},
  {Frazer}, {Lanning}, {Misch}, {Mueller}, {Soyumer}, {Woodard}, {Baliunas},
  {Noyes}, {Hartmann}, {Porter}, {Zwaan}, {Middelkoop}, {Rutten}, \&
  {Mihalas}}]{duncan1991}
{Duncan}, D.~K., {Vaughan}, A.~H., {Wilson}, O.~C., {et~al.} 1991, \apjs, 76,
  383

\bibitem[{{Fabrycky} \& {Tremaine}(2007)}]{Fabrycky07}
{Fabrycky}, D., \& {Tremaine}, S. 2007, \apj, 669, 1298

\bibitem[{{Feinstein} {et~al.}(2019){Feinstein}, {Montet}, {Foreman-Mackey},
  {Bedell}, {Saunders}, {Bean}, {Christiansen}, {Hedges}, {Luger}, {Scolnic},
  \& {Cardoso}}]{Feinstein19}
{Feinstein}, A.~D., {Montet}, B.~T., {Foreman-Mackey}, D., {et~al.} 2019,
  \pasp, 131, 094502

\bibitem[{{Ford}(2014)}]{Ford14}
{Ford}, E.~B. 2014, Proceedings of the National Academy of Science, 111, 12616

\bibitem[{{Foreman-Mackey} {et~al.}(2013){Foreman-Mackey}, {Hogg}, {Lang}, \&
  {Goodman}}]{Foreman-Mackey12}
{Foreman-Mackey}, D., {Hogg}, D.~W., {Lang}, D., \& {Goodman}, J. 2013, \pasp,
  125, 306

\bibitem[{{Franchini} {et~al.}(2020){Franchini}, {Martin}, \&
  {Lubow}}]{Franchini19}
{Franchini}, A., {Martin}, R.~G., \& {Lubow}, S.~H. 2020, \mnras, 491, 5351

\bibitem[{{Fr{\"o}hlich} {et~al.}(2012){Fr{\"o}hlich}, {Frasca}, {Catanzaro},
  {Bonanno}, {Corsaro}, {Molenda-{\.Z}akowicz}, {Klutsch}, \&
  {Montes}}]{Frohlich12}
{Fr{\"o}hlich}, H.~E., {Frasca}, A., {Catanzaro}, G., {et~al.} 2012, \aap, 543,
  A146

\bibitem[{{Gandolfi} {et~al.}(2010){Gandolfi}, {H{\'e}brard}, {Alonso},
  {Deleuil}, {Guenther}, {Fridlund}, {Endl}, {Eigm{\"u}ller}, {Csizmadia},
  {Havel}, {Aigrain}, {Auvergne}, {Baglin}, {Barge}, {Bonomo}, {Bord{\'e}},
  {Bouchy}, {Bruntt}, {Cabrera}, {Carpano}, {Carone}, {Cochran}, {Deeg},
  {Dvorak}, {Eisl{\"o}ffel}, {Erikson}, {Ferraz-Mello}, {Gazzano}, {Gibson},
  {Gillon}, {Gondoin}, {Guillot}, {Hartmann}, {Hatzes}, {Jorda}, {Kabath},
  {L{\'e}ger}, {Llebaria}, {Lammer}, {MacQueen}, {Mayor}, {Mazeh}, {Moutou},
  {Ollivier}, {P{\"a}tzold}, {Pepe}, {Queloz}, {Rauer}, {Rouan}, {Samuel},
  {Schneider}, {Stecklum}, {Tingley}, {Udry}, \& {Wuchterl}}]{Gandolfi10}
{Gandolfi}, D., {H{\'e}brard}, G., {Alonso}, R., {et~al.} 2010, \aap, 524, A55

\bibitem[{{Geweke}(1992)}]{Geweke92}
{Geweke}, J. 1992, Bayesian Statistics IV. Oxford: Clarendon Press, ed. J.~M.
  {Bernardo}, 169

\bibitem[{{Gim{\'e}nez}(2006)}]{Gimenez06}
{Gim{\'e}nez}, A. 2006, \apj, 650, 408

\bibitem[{{Gomes da Silva} {et~al.}(2011){Gomes da Silva}, {Santos}, {Bonfils},
  {Delfosse}, {Forveille}, \& {Udry}}]{gomesdasilva2011}
{Gomes da Silva}, J., {Santos}, N.~C., {Bonfils}, X., {et~al.} 2011, \aap, 534,
  A30

\bibitem[{Goodman \& Weare(2010)}]{Goodman10}
Goodman, J., \& Weare, J. 2010, Communications in Applied Mathematics and
  Computational Science, 5, 65

\bibitem[{{Hirano} {et~al.}(2010){Hirano}, {Suto}, {Taruya}, {Narita}, {Sato},
  {Johnson}, \& {Winn}}]{Hirano10}
{Hirano}, T., {Suto}, Y., {Taruya}, A., {et~al.} 2010, \apj, 709, 458

\bibitem[{{Hirano} {et~al.}(2012){Hirano}, {Narita}, {Sato}, {Takahashi},
  {Masuda}, {Takeda}, {Aoki}, {Tamura}, \& {Suto}}]{Hirano12b}
{Hirano}, T., {Narita}, N., {Sato}, B., {et~al.} 2012, \apjl, 759, L36

\bibitem[{{Holman} {et~al.}(1997){Holman}, {Touma}, \& {Tremaine}}]{Holman97}
{Holman}, M., {Touma}, J., \& {Tremaine}, S. 1997, \nat, 386, 254

\bibitem[{Hunter {et~al.}(2007)}]{matplotlib}
Hunter, J.~D., {et~al.} 2007, Computing in science and engineering, 9, 90

\bibitem[{{Ida} \& {Lin}(2008)}]{Ida08}
{Ida}, S., \& {Lin}, D.~N.~C. 2008, \apj, 673, 487

\bibitem[{{Jenkins} {et~al.}(2010){Jenkins}, {Borucki}, {Koch}, {Marcy},
  {Cochran}, {Welsh}, {Basri}, {Batalha}, {Buchhave}, {Brown}, {Caldwell},
  {Dunham}, {Endl}, {Fischer}, {Gautier}, {Geary}, {Gilliland}, {Howell},
  {Isaacson}, {Johnson}, {Latham}, {Lissauer}, {Monet}, {Rowe}, {Sasselov},
  {Howard}, {MacQueen}, {Orosz}, {Chandrasekaran}, {Twicken}, {Bryson},
  {Quintana}, {Clarke}, {Li}, {Allen}, {Tenenbaum}, {Wu}, {Meibom}, {Klaus},
  {Middour}, {Cote}, {McCauliff}, {Girouard}, {Gunter}, {Wohler}, {Hall},
  {Ibrahim}, {Kamal Uddin}, {Wu}, {Bhavsar}, {Van Cleve}, {Pletcher}, {Dotson},
  \& {Haas}}]{Jenkins10b}
{Jenkins}, J.~M., {Borucki}, W.~J., {Koch}, D.~G., {et~al.} 2010, \apj, 724,
  1108

\bibitem[{{Johnson} {et~al.}(2008){Johnson}, {Winn}, {Narita}, {Enya},
  {Williams}, {Marcy}, {Sato}, {Ohta}, {Taruya}, {Suto}, {Turner}, {Bakos},
  {Butler}, {Vogt}, {Aoki}, {Tamura}, {Yamada}, {Yoshii}, \&
  {Hidas}}]{Johnson08}
{Johnson}, J.~A., {Winn}, J.~N., {Narita}, N., {et~al.} 2008, \apj, 686, 649

\bibitem[{Jones {et~al.}(2001--)Jones, Oliphant, Peterson, {et~al.}}]{Jones01}
Jones, E., Oliphant, T., Peterson, P., {et~al.} 2001--, {SciPy}: Open source
  scientific tools for {Python}, , .
\newblock \url{http://www.scipy.org/}

\bibitem[{{Kilcik} {et~al.}(2011){Kilcik}, {Yurchyshyn}, {Abramenko}, {Goode},
  {Ozguc}, {Rozelot}, \& {Cao}}]{Kilcik11}
{Kilcik}, A., {Yurchyshyn}, V.~B., {Abramenko}, V., {et~al.} 2011, \apj, 731,
  30

\bibitem[{{Kipping}(2013)}]{Kipping13b}
{Kipping}, D.~M. 2013, \mnras, 435, 2152

\bibitem[{{Kochanek} {et~al.}(2017){Kochanek}, {Shappee}, {Stanek}, {Holoien},
  {Thompson}, {Prieto}, {Dong}, {Shields}, {Will}, {Britt}, {Perzanowski}, \&
  {Pojma{\'n}ski}}]{Kochanek17}
{Kochanek}, C.~S., {Shappee}, B.~J., {Stanek}, K.~Z., {et~al.} 2017, \pasp,
  129, 104502

\bibitem[{{Kozai}(1962)}]{Kozai62}
{Kozai}, Y. 1962, \aj, 67, 591

\bibitem[{{Lanza} {et~al.}(2019){Lanza}, {Collier Cameron}, \&
  {Haywood}}]{Lanza19}
{Lanza}, A.~F., {Collier Cameron}, A., \& {Haywood}, R.~D. 2019, \mnras, 486,
  3459

\bibitem[{{Lidov}(1962)}]{Lidov62}
{Lidov}, M.~L. 1962, \planss, 9, 719

\bibitem[{{Liu} {et~al.}(2002){Liu}, {Fischer}, {Graham}, {Lloyd}, {Marcy}, \&
  {Butler}}]{Liu02}
{Liu}, M.~C., {Fischer}, D.~A., {Graham}, J.~R., {et~al.} 2002, \apj, 571, 519

\bibitem[{{Luger} {et~al.}(2019){Luger}, {Agol}, {Foreman-Mackey}, {Fleming},
  {Lustig-Yaeger}, \& {Deitrick}}]{Luger19}
{Luger}, R., {Agol}, E., {Foreman-Mackey}, D., {et~al.} 2019, \aj, 157, 64

\bibitem[{{Mann} {et~al.}(2016){Mann}, {Newton}, {Rizzuto}, {Irwin}, {Feiden},
  {Gaidos}, {Mace}, {Kraus}, {James}, {Ansdell}, {Charbonneau}, {Covey},
  {Ireland}, {Jaffe}, {Johnson}, {Kidder}, \& {Vanderburg}}]{Mann16}
{Mann}, A.~W., {Newton}, E.~R., {Rizzuto}, A.~C., {et~al.} 2016, \aj, 152, 61

\bibitem[{{Marcy} \& {Butler}(1992)}]{Marcy92}
{Marcy}, G.~W., \& {Butler}, R.~P. 1992, \pasp, 104, 270

\bibitem[{McLachlan \& Peel(2000)}]{McLachlan00}
McLachlan, G.~J., \& Peel, D. 2000, Finite mixture models (New York: Wiley
  Series in Probability and Statistics)

\bibitem[{{McLaughlin}(1924)}]{McLaughlin24}
{McLaughlin}, D.~B. 1924, \apj, 60, 22

\bibitem[{{McQuillan} {et~al.}(2014){McQuillan}, {Mazeh}, \&
  {Aigrain}}]{McQuillan14}
{McQuillan}, A., {Mazeh}, T., \& {Aigrain}, S. 2014, \apjs, 211, 24

\bibitem[{{Meunier} {et~al.}(2019){Meunier}, {Lagrange}, \&
  {Cuzacq}}]{Meunier19}
{Meunier}, N., {Lagrange}, A.~M., \& {Cuzacq}, S. 2019, \aap, 632, A81

\bibitem[{{Millholland} \& {Laughlin}(2019)}]{Millholland19}
{Millholland}, S., \& {Laughlin}, G. 2019, Nature Astronomy, 3, 424

\bibitem[{{Montet} {et~al.}(2017){Montet}, {Tovar}, \&
  {Foreman-Mackey}}]{Montet17}
{Montet}, B.~T., {Tovar}, G., \& {Foreman-Mackey}, D. 2017, \apj, 851, 116

\bibitem[{{Montet} {et~al.}(2015){Montet}, {Bowler}, {Shkolnik}, {Deck},
  {Wang}, {Horch}, {Liu}, {Hillenbrand}, {Kraus}, \& {Charbonneau}}]{Montet15c}
{Montet}, B.~T., {Bowler}, B.~P., {Shkolnik}, E.~L., {et~al.} 2015, \apjl, 813,
  L11

\bibitem[{{Morris} {et~al.}(2017){Morris}, {Hebb}, {Davenport}, {Rohn}, \&
  {Hawley}}]{Morris17}
{Morris}, B.~M., {Hebb}, L., {Davenport}, J. R.~A., {Rohn}, G., \& {Hawley},
  S.~L. 2017, \apj, 846, 99

\bibitem[{{Morton} \& {Johnson}(2011)}]{Morton11a}
{Morton}, T.~D., \& {Johnson}, J.~A. 2011, \apj, 729, 138

\bibitem[{{Naoz}(2016)}]{Naoz16}
{Naoz}, S. 2016, \araa, 54, 441

\bibitem[{{Nelson} {et~al.}(2017){Nelson}, {Ford}, \& {Rasio}}]{Nelson17}
{Nelson}, B.~E., {Ford}, E.~B., \& {Rasio}, F.~A. 2017, \aj, 154, 106

\bibitem[{{Neveu-VanMalle} {et~al.}(2016){Neveu-VanMalle}, {Queloz},
  {Anderson}, {Brown}, {Collier Cameron}, {Delrez}, {D{\'\i}az}, {Gillon},
  {Hellier}, {Jehin}, {Lister}, {Pepe}, {Rojo}, {S{\'e}gransan}, {Triaud},
  {Turner}, \& {Udry}}]{NeveuVanMalle16}
{Neveu-VanMalle}, M., {Queloz}, D., {Anderson}, D.~R., {et~al.} 2016, \aap,
  586, A93

\bibitem[{{Newton} {et~al.}(2019){Newton}, {Mann}, {Tofflemire}, {Pearce},
  {Rizzuto}, {Vanderburg}, {Martinez}, {Wang}, {Ruffio}, {Kraus}, {Johnson},
  {Thao}, {Wood}, {Rampalli}, {Nielsen}, {Collins}, {Dragomir}, {Hellier},
  {Anderson}, {Barclay}, {Brown}, {Feiden}, {Hart}, {Isopi}, {Kielkopf},
  {Mallia}, {Nelson}, {Rodriguez}, {Stockdale}, {Waite}, {Wright}, {Lissauer},
  {Ricker}, {Vanderspek}, {Latham}, {Seager}, {Winn}, {Jenkins}, {Bouma},
  {Burke}, {Davies}, {Fausnaugh}, {Li}, {Morris}, {Mukai}, {Villase{\~n}or},
  {Villeneuva}, {De Rosa}, {Macintosh}, {Mengel}, {Okumura}, \&
  {Wittenmyer}}]{Newton19}
{Newton}, E.~R., {Mann}, A.~W., {Tofflemire}, B.~M., {et~al.} 2019, \apjl, 880,
  L17

\bibitem[{{Oshagh} {et~al.}(2018){Oshagh}, {Triaud}, {Burdanov}, {Figueira},
  {Reiners}, {Santos}, {Faria}, {Boue}, {D{\'\i}az}, {Dreizler}, {Boldt},
  {Delrez}, {Ducrot}, {Gillon}, {Guzman Mesa}, {Jehin}, {Khalafinejad}, {Kohl},
  {Serrano}, \& {Udry}}]{Oshagh18}
{Oshagh}, M., {Triaud}, A.~H.~M.~J., {Burdanov}, A., {et~al.} 2018, \aap, 619,
  A150

\bibitem[{{Reiners}(2009)}]{Reiners09}
{Reiners}, A. 2009, \aap, 498, 853

\bibitem[{{Ricker} {et~al.}(2014){Ricker}, {Winn}, {Vanderspek}, {Latham},
  {Bakos}, {Bean}, {Berta-Thompson}, {Brown}, {Buchhave}, {Butler}, {Butler},
  {Chaplin}, {Charbonneau}, {Christensen-Dalsgaard}, {Clampin}, {Deming},
  {Doty}, {De Lee}, {Dressing}, {Dunham}, {Endl}, {Fressin}, {Ge}, {Henning},
  {Holman}, {Howard}, {Ida}, {Jenkins}, {Jernigan}, {Johnson}, {Kaltenegger},
  {Kawai}, {Kjeldsen}, {Laughlin}, {Levine}, {Lin}, {Lissauer}, {MacQueen},
  {Marcy}, {McCullough}, {Morton}, {Narita}, {Paegert}, {Palle}, {Pepe},
  {Pepper}, {Quirrenbach}, {Rinehart}, {Sasselov}, {Sato}, {Seager},
  {Sozzetti}, {Stassun}, {Sullivan}, {Szentgyorgyi}, {Torres}, {Udry}, \&
  {Villasenor}}]{Ricker14}
{Ricker}, G.~R., {Winn}, J.~N., {Vanderspek}, R., {et~al.} 2014, in \procspie,
  Vol. 9143, Space Telescopes and Instrumentation 2014: Optical, Infrared, and
  Millimeter Wave, 914320

\bibitem[{{Robertson} {et~al.}(2013){Robertson}, {Endl}, {Cochran}, \&
  {Dodson-Robinson}}]{robertson13}
{Robertson}, P., {Endl}, M., {Cochran}, W.~D., \& {Dodson-Robinson}, S.~E.
  2013, \apj, 764, 3

\bibitem[{{Robertson} {et~al.}(2015){Robertson}, {Roy}, \&
  {Mahadevan}}]{Robertson15}
{Robertson}, P., {Roy}, A., \& {Mahadevan}, S. 2015, \apjl, 805, L22

\bibitem[{{Rossiter}(1924)}]{Rossiter24}
{Rossiter}, R.~A. 1924, \apj, 60, 15

\bibitem[{{Sanchis-Ojeda} {et~al.}(2013){Sanchis-Ojeda}, {Winn}, {Marcy},
  {Howard}, {Isaacson}, {Johnson}, {Torres}, {Albrecht}, {Campante}, {Chaplin},
  {Davies}, {Lund}, {Carter}, {Dawson}, {Buchhave}, {Everett}, {Fischer},
  {Geary}, {Gilliland}, {Horch}, {Howell}, \& {Latham}}]{SanchisOjeda13}
{Sanchis-Ojeda}, R., {Winn}, J.~N., {Marcy}, G.~W., {et~al.} 2013, \apj, 775,
  54

\bibitem[{{Santos} {et~al.}(2000){Santos}, {Mayor}, {Naef}, {Pepe}, {Queloz},
  {Udry}, \& {Blecha}}]{santos2000}
{Santos}, N.~C., {Mayor}, M., {Naef}, D., {et~al.} 2000, \aap, 361, 265

\bibitem[{{Shappee} {et~al.}(2014){Shappee}, {Prieto}, {Grupe}, {Kochanek},
  {Stanek}, {De Rosa}, {Mathur}, {Zu}, {Peterson}, {Pogge}, {Komossa}, {Im},
  {Jencson}, {Holoien}, {Basu}, {Beacom}, {Szczygie{\l}}, {Brimacombe},
  {Adams}, {Campillay}, {Choi}, {Contreras}, {Dietrich}, {Dubberley},
  {Elphick}, {Foale}, {Giustini}, {Gonzalez}, {Hawkins}, {Howell}, {Hsiao},
  {Koss}, {Leighly}, {Morrell}, {Mudd}, {Mullins}, {Nugent}, {Parrent},
  {Phillips}, {Pojmanski}, {Rosing}, {Ross}, {Sand}, {Terndrup}, {Valenti},
  {Walker}, \& {Yoon}}]{Shappee14}
{Shappee}, B.~J., {Prieto}, J.~L., {Grupe}, D., {et~al.} 2014, \apj, 788, 48

\bibitem[{{Short} {et~al.}(2018){Short}, {Orosz}, {Windmiller}, \&
  {Welsh}}]{Short18}
{Short}, D.~R., {Orosz}, J.~A., {Windmiller}, G., \& {Welsh}, W.~F. 2018, \aj,
  156, 297

\bibitem[{{Soderblom}(2010)}]{Soderblom10}
{Soderblom}, D.~R. 2010, \araa, 48, 581

\bibitem[{{Triaud} {et~al.}(2009){Triaud}, {Queloz}, {Bouchy}, {Moutou},
  {Collier Cameron}, {Claret}, {Barge}, {Benz}, {Deleuil}, {Guillot},
  {H{\'e}brard}, {Lecavelier Des {\'E}tangs}, {Lovis}, {Mayor}, {Pepe}, \&
  {Udry}}]{Triaud09}
{Triaud}, A.~H.~M.~J., {Queloz}, D., {Bouchy}, F., {et~al.} 2009, \aap, 506,
  377

\bibitem[{Van Der~Walt {et~al.}(2011)Van Der~Walt, Colbert, \&
  Varoquaux}]{numpy}
Van Der~Walt, S., Colbert, S.~C., \& Varoquaux, G. 2011, Computing in Science
  \& Engineering, 13, 22

\bibitem[{{Waite} {et~al.}(2017){Waite}, {Marsden}, {Carter}, {Petit},
  {Jeffers}, {Morin}, {Vidotto}, {Donati}, \& {BCool Collaboration}}]{Waite17}
{Waite}, I.~A., {Marsden}, S.~C., {Carter}, B.~D., {et~al.} 2017, \mnras, 465,
  2076

\bibitem[{{Winn} {et~al.}(2005){Winn}, {Noyes}, {Holman}, {Charbonneau},
  {Ohta}, {Taruya}, {Suto}, {Narita}, {Turner}, {Johnson}, {Marcy}, {Butler},
  \& {Vogt}}]{Winn05}
{Winn}, J.~N., {Noyes}, R.~W., {Holman}, M.~J., {et~al.} 2005, \apj, 631, 1215

\bibitem[{{Winn} {et~al.}(2008){Winn}, {Johnson}, {Narita}, {Suto}, {Turner},
  {Fischer}, {Butler}, {Vogt}, {O'Donovan}, \& {Gaudi}}]{Winn08}
{Winn}, J.~N., {Johnson}, J.~A., {Narita}, N., {et~al.} 2008, \apj, 682, 1283

\bibitem[{{Zanazzi} \& {Lai}(2018)}]{Zanazzi18}
{Zanazzi}, J.~J., \& {Lai}, D. 2018, \mnras, 478, 835

\bibitem[{{Zhou} {et~al.}(2019){Zhou}, {Winn}, {Newton}, {Quinn}, {Rodriguez},
  {Mann}, {Rizzuto}, {Vand erburg}, {Huang}, {Latham}, {Teske}, {Wang},
  {Shectman}, {Butler}, {Crane}, {Thompson}, {Henry}, {Paredes}, {Jao},
  {James}, \& {Hinojosa}}]{Zhou19}
{Zhou}, G., {Winn}, J.~N., {Newton}, E.~R., {et~al.} 2019, arXiv e-prints,
  arXiv:1912.04095

\end{thebibliography}
\end{document}